\documentclass[notitlepage]{revtex4} 
\usepackage{graphicx}

\usepackage{hyperref} 					 
\usepackage{listings} 					 
\usepackage{color}    					 
\usepackage{natbib}   					 
\usepackage{subfig} 					 

\usepackage{booktabs}
\usepackage[ansinew]{inputenc}				     
\usepackage{psfrag}
\usepackage{pstricks}
\usepackage{amsfonts}
\usepackage{amsmath}
\usepackage{bm}
\usepackage{lipsum}
\usepackage{pgf}
\usepackage{tikz}
\usetikzlibrary{patterns}

\usepackage{fancyvrb}
\usepackage{listings}
\usepackage{color} 
\definecolor{mygreen}{RGB}{28,172,0} 
\definecolor{mylilas}{RGB}{170,55,241}
\usepackage{url} 
\usepackage[normalem]{ulem} 
\useunder{\uline}{\ul}{} 

\usepackage{array}
\newcolumntype{C}[1]{>{\centering\arraybackslash}m{#1}}

\usepackage{algpseudocode}
\usepackage{color}

\usepackage{filecontents}

\begin{document}	

\title{Inverse design in photonics by topology optimization: tutorial}

\author{Rasmus E. Christiansen,$^{1,2}$}
\email{raelch@mek.dtu.dk}

\author{Ole Sigmund,$^{1,2}$}

\affiliation{$^{1}$ Department of Mechanical Engineering, Technical University of Denmark, Nils Koppels All\'{e}, Building 404, 2800 Kongens Lyngby, Denmark}
\affiliation{$^{2}$ NanoPhoton---Center for Nanophotonics, Technical University of Denmark, {\O}rsteds Plads 345A, DK-2800 Kgs. Lyngby, Denmark.}

\begin{abstract}
	Topology optimization methods for inverse design of nano-photonic systems have recently become extremely popular and are presented in various forms and under various names. Approaches comprise gradient and non-gradient based algorithms combined with more or less systematic ways to improve convergence, discreteness of solutions and satisfaction of manufacturing constraints. We here provide a tutorial for the systematic and efficient design of nano-photonic structures by Topology Optimization (TopOpt). The implementation is based on the advanced and systematic approaches developed in TopOpt for structural optimization during the last three decades. The tutorial presents a step-by-step guide for deriving the continuous constrained optimization problem forming the foundation of the Topology Optimization method, using a cylindrical metalens design problem as an example. It demonstrates the effect and necessity of applying a number of auxiliary tools in the design process in order to ensure good numerical modelling practice and to achieve physically realisable designs. Application examples also include an optical demultiplexer.
\end{abstract}

\maketitle

\section{Introduction}

We provide an introduction and tutorial for using density-based Topology Optimization (TopOpt) as an inverse design tool for photonic structures. We use the commercial software package COMSOL Multiphysics \cite{COMSOL55} for the numerical implementation, and provide a set of COMSOL models alongside the article (available on \url{https://www.topopt.mek.dtu.dk}). These models allow for replication of the presented results at the click of a button, as well as providing a starting point for using TopOpt for more advanced photonics applications. After an initial definition of the spatial, temporal and physics model (Secs.~\ref{SEC:SPACE}-\ref{SEC:PHYSICS}), we demonstrate how to derive the basic continuous constrained optimization problem, which forms the foundation for using TopOpt for inverse design, using a cylindrical metalens design problem as an example (Sec.~\ref{SEC:OPTIMIZATION_PROBLEM}). The basic optimization problem derived in Sec.~\ref{SEC:OPTIMIZATION_PROBLEM} is solved to illustrate how this first naive approach leads to non-physical solutions (Sec.~\ref{SEC:METALENS_CASE_1}). Following this, the problem formulation is extended (step-by-step Sec.~\ref{SEC:METALENS_CASE_2}-\ref{SEC:METALENS_CASE_3}), until it ensures that solutions are physically sensible and support the design of multi-wavelength metalenses (Sec.~\ref{SEC:METALENS_CASE_4}). As a second example we consider an optical demultiplexer, demonstrating how the TopOpt problem can be extended to ensure that the solution exhibits robustness towards near-uniform geometric perturbations (Sec.~\ref{SEC:BEAMSPLITTER}). Finally, we give a brief introduction to more advanced TopOpt tools, which among others, include tools for imposing geometric length-scales in the design; for ensuring the connectivity of the design; and for ensuring that the design conforms to manufacturing constraints (Sec.~\ref{SEC:ADVANCED}).

Stated succinctly, density-based Topology Optimization \cite{BENDSOE_KIKUCHI_1988,BOOK_TOPOPT_BENDSOE} is an inverse design tool, used to produce highly optimized structures to serve specialized purposes, applicable across most areas of physics \cite{AAGE_ET_AL2017,ALEXANDERSEN_ANDREASEN_2020,LUNDGAARD_2018,JENSEN_SIGMUND_2011,Christiansen_Grande_2016}. A defining feature is that density-based TopOpt uses adjoint analysis for efficient gradient computations. Conceptually TopOpt offers unparalleled design freedom, as it allows for point-by-point variation in the material distribution constituting the structure under design, with recent work demonstrating the solution of design problems with more than a billion design variables \textcolor{black}{for mechanical problems} \cite{AAGE_ET_AL2017}, proving that TopOpt in any practical sense is able to provide unlimited design freedom. Indeed, the main challenge when applying TopOpt is often to limit the design freedom offered by the method, in a way that conforms with fabrication constraints. Within photonics and plasmonics, TopOpt has received increasing attention over the last two decades \cite{JENSEN_SIGMUND_2011,MOLESKY_2018} with recent examples of applicationss including the design of dielectric multiplexers \cite{PIGGOTT_2015}, dielectric metalenses \cite{ZIN_2019,CHUNG_MILLER_2020,CHRISTIANSEN_2020b}, extreme dielectric confinement structures \textcolor{black}{\cite{LIANG_JOHNSON_2013,WANG_2018}}, plasmonic nano-antennas \cite{WADBRO_ENGSTROM_2015}, plasmonic nanoparticles for solar cell applications \cite{MADSEN_2020,CHRISTIANSEN_SEMSC_2020}, for enhanced thermal emission \cite{KUDYSHEV_ET_AL_2020} and Raman scattering \cite{CHRISTIANSEN_OE_2020}, to name but a few. 

\textcolor{black}{By now, inverse-design methods go by many names. The basic topology-optimization concepts originate from the structural-optimization community, whereas the photonics community has adopted related inverse design approaches \cite{MOLESKY_2018} and so-called objective first approaches \cite{LU_VUCKOVIC_2012}, which mainly differ in the way the optimization problems are defined and solved.} An indispensable part of all inverse-design tools with many degrees of freedom is the adjoint sensitivity analysis \cite{JENSEN_SIGMUND_2011}. 

In this work we consider electromagnetism modelled using Maxwell's equations assuming linear, static, homogeneous, isotropic, non-dispersive and non-magnetic materials. We assume time harmonic behaviour of the field and only consider transverse electric and transverse magnetic problems with material invariance in the polarization direction. All of these assumptions are made for simplicity and are not required in order to utilize TopOpt in the context of electromagnetism. 

For readers who are interested in \textcolor{black}{underlying} method development, programming and software implementation, we have authored a parallel tutorial paper describing a freely available 200 line MATLAB code implementing basic TopOpt for photonics~\cite{CHRISTIANSEN_SIGMUND_MATLAB_2020}. \textcolor{black}{That paper also includes an example demonstrating the advantages of gradient-based methods over so-called global-optimization methods for the type of inverse design problems considered here.} 

\section{The Goal} \label{SEC:GOAL}

The ultimate goal in any structural design process is to identify the structure that best solves the problem at hand. A more operative formulation of this goal is: 

\begin{center}
	\textit{For a given structural design problem the goal is to identify a structure that maximizes the desired figure(s) of merit, without violating any of the constraints imposed on the problem.}
\end{center}
	
\section{Space} \label{SEC:SPACE}

\begin{figure}[h!]
	\centering
	{
		\includegraphics[width=0.7\textwidth]{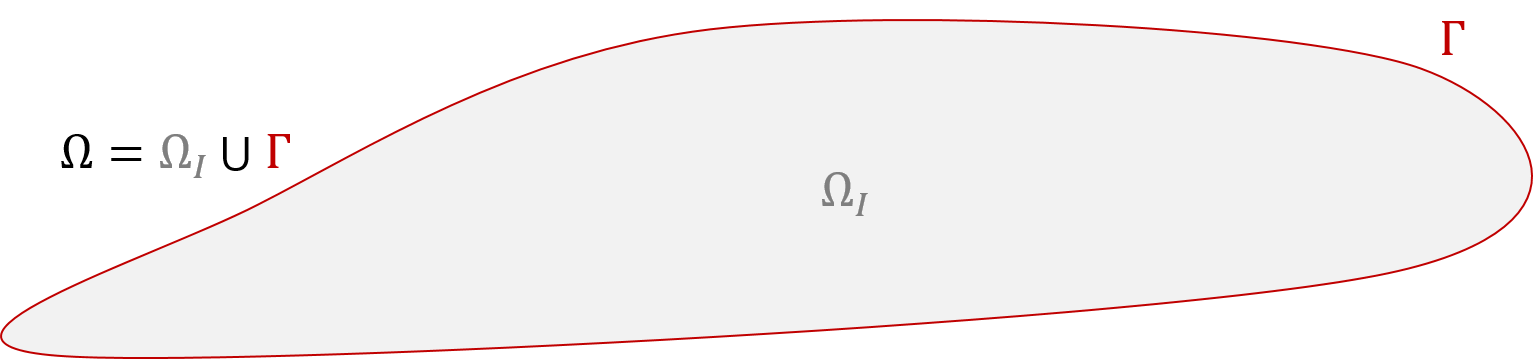} \caption{Sketch of model domain $\Omega = \Omega_{I} \bigcup \Gamma$ in 2D, $\Omega_{I}$ denotes the interior and $\Gamma$ the boundary. \label{FIG:MODEL_DOMAIN}}
	}
\end{figure}

We assume a Cartesian coordinate system to describe space, i.e. $\textbf{r} = \lbrace x,y,z \rbrace \in \mathbb{R}^3$ in three dimensions and $\textbf{r} = \lbrace x,y \rbrace \in \mathbb{R}^2$ in two dimensions, where $\mathbb{R}$ denotes the real numbers. For numerical modelling of the physics we define a spatially limited modelling domain, $\Omega$, with the interior $\Omega_{I}$ and the boundary $\Gamma$, as illustrated in figure \ref{FIG:MODEL_DOMAIN}.

\section{Time} \label{SEC:TIME}

We consider problems which are time-harmonic in nature and disregard any transient behaviour. The time dependence is therefore modelled using the time-harmonic exponential factor, $e^{\mathrm{i}\omega t}$, where $t$ denotes time, $\omega$ the angular frequency and $\mathrm{i}$ the imaginary unit.

\section{Physics} \label{SEC:PHYSICS}

Under the aforementioned assumptions, we consider the following field equations for the electric field, $\mathcal{E}$, and magnetic field, $\mathcal{H} = \frac{1}{\mu_0} \mathcal{B}$ \cite{BOOK_EM_GRIFFITHS},

\begin{eqnarray}
\nabla \cdot \mathcal{E} = \frac{\rho}{\varepsilon_r \varepsilon_0}, \ \
\nabla \cdot \mathcal{H} = 0, \ \
\nabla \times \mathcal{E} = -\mu_0\frac{\partial \mathcal{H}}{\partial t}, \ \
\nabla \times \mathcal{H} = \textbf{J}_{\mathrm{f}}+\varepsilon_r \varepsilon_0\frac{\partial \mathcal{E}}{\partial t}, \ \ \mathcal{E} = \textbf{E} e^{\mathrm{i} \omega t}, \ \ \ \mathcal{H} = \textbf{H} e^{\mathrm{i} \omega t},  \label{EQN:MAXWELL_FULL}
\end{eqnarray}

\noindent Here $\textbf{J}_{\mathrm{f}}$ and $\rho$ denote the free-current and free-charge densities, and $\varepsilon_0$ and $\mu_0$ denote the vacuum electric permittivity and the vacuum magnetic permeability, respectively. Further, $\varepsilon_r$ denotes the relative electric permittivity of the medium through which $\mathcal{E}$ and $\mathcal{H}$ propagate and finally $\textbf{E}$ and $\textbf{H}$ denote the spatially dependent part of the electric and magnetic fields.

We assume that the current and charge densities are zero in the interior of the model domain, i.e. $\textbf{J}_{\mathrm{f}}(\textbf{r}) = \textbf{0}, \ \rho(\textbf{r}) = 0, \ \ \textbf{r} \in \boldmath{\Omega}_I$. Thereby, the following equations for $\textbf{E}$ and $\textbf{H}$ in $\Omega_{I}$ are derived,
 
\begin{eqnarray}
\nabla \times \nabla \times \textbf{E}(\textbf{r}) - \frac{\omega^2}{c^2} \varepsilon_r(\textbf{r}) \textbf{E}(\textbf{r}) = \textbf{0}, \ \ \ \textbf{r} \in \boldmath{\Omega}_I \subset \mathbb{R}^3, \label{EQN:MAXWELL_EQUATION_E_FREQUENCY_DOMAIN} \\ 
\nabla \times \left( \frac{1}{\varepsilon_r(\textbf{r})}  \nabla \times \textbf{H}(\textbf{r}) \right) - \frac{\omega^2}{c^2} \textbf{H}(\textbf{r}) = \textbf{0}, \ \ \ \textbf{r} \in \boldmath{\Omega}_I \subset \mathbb{R}^3. \label{EQN:MAXWELL_EQUATION_B_FREQUENCY_DOMAIN} 
\end{eqnarray}

\noindent Here $c = 1/ \sqrt{\mu_0 \varepsilon_0}$ denotes the speed of light in vacuum. In addition to eqs.~(\ref{EQN:MAXWELL_EQUATION_E_FREQUENCY_DOMAIN})-(\ref{EQN:MAXWELL_EQUATION_B_FREQUENCY_DOMAIN}), problem specific boundary conditions are imposed on the boundary of the model domain, $\Gamma$, in order to truncate it appropriately and to introduce external fields.

\subsection{Two Dimensional Model}

Assuming material invariance in the out-of-plane direction (the $z$-direction) and assuming that either $\textbf{E}$ or $\textbf{H}$ is linearly polarized in the $z$-direction allows for the reduction of  eq.~(\ref{EQN:MAXWELL_EQUATION_E_FREQUENCY_DOMAIN})[eq.~(\ref{EQN:MAXWELL_EQUATION_B_FREQUENCY_DOMAIN})] to a scalar Helmholtz equation in two dimensions given as,

\begin{eqnarray}
\mathcal{L}_{EM}(\phi) = \nabla \cdot (a \ \nabla \phi) + \frac{\omega^2}{c^2} b \ \phi = 0, \ \ \textbf{r} \in \boldmath{\Omega}_I \subset \mathbb{R}^2.  \label{EQN:EM_HELMHOLTZ_2D}
\end{eqnarray}

To model a problem for an $E_z$-polarized field ($E_x = E_y = 0$), denoted TE in the following, one selects $\phi = E_z$, $a = 1$ and $b = \varepsilon_r$. To model a problem for an $H_z$-polarized field ($H_x = H_y = 0$), denoted TM in the following, one selects $\phi = H_z$, $a = 1 / \varepsilon_r$ and $b = 1$. From the solution of eq.~(\ref{EQN:EM_HELMHOLTZ_2D}) $\mathcal{E}$ and $\mathcal{H}$ ($\textbf{E}$ and $\textbf{H}$) may be computed using eq.~(\ref{EQN:MAXWELL_FULL}).

\section{Optimization Problem} \label{SEC:OPTIMIZATION_PROBLEM}

\noindent \textit{As an example of the derivation of the optimization problem forming the basis of TopOpt, we consider the design problem treated in Sec.~\ref{SEC:METALENS_CASE_1}, which may be stated informally as: } 

\begin{center}
 \textit{\textbf{Design a cylindrical\footnote{A cylindrical lens focuses the incident field into a line rather than a point.} silicon metalens capable of monochromatic focusing of TM-polarized light at normal incidence into a focal line behind the lens.}}  
\end{center}

\noindent \textit{Neglecting the field behaviour at the lens ends, this problem may be modelled in 2D by assuming material invariance in the out-of-plane direction, effectively turning the focal line in 3D into a focal point in 2D.} \\

\noindent To solve any structural design problem using TopOpt, it must be formulated as a continuous constrained optimization problem, formally written as,

\begin{eqnarray}
\underset{\xi}{\max} &\Phi(\xi),& \ \ \Phi:~\left[0,1\right]^{\Omega_{d}}\rightarrow\mathbb{R}, \nonumber \\
\mathrm{s.t.} &c_i(\xi) = 0,& \ \  c_i:~\left[0,1\right]^{\Omega_{d}}\rightarrow\mathbb{R}, \ \ i \in \lbrace 0,1, ..., \mathcal{N}_{i} \rbrace, \ \ \mathcal{N}_{i} \in \mathbb{N}_0 \label{EQN:OPTIMIZATION_PROBLEM_EXAMPLE}  \\
 &c_j(\xi) < 0,& \ \   c_j:~\left[0,1\right]^{\Omega_{d}}\rightarrow\mathbb{R}, \ \ j \in \lbrace 0,1, ..., \mathcal{N}_{j} \rbrace, \ \ \mathcal{N}_{j} \in \mathbb{N}_0. \nonumber
\end{eqnarray}

Here $\xi(\textbf{r}) \in [0,1]$ denotes a continuous field, called \textit{the design field}, over which the function $\Phi$, called \textit{the figure of merit} (FOM), is sought maximized. The equations $c_i=0$ denote $\mathcal{N}_{i}$ equality constraints and the inequalities $c_j<0$ denote $\mathcal{N}_{j}$ inequality constraints. Looking at eq.~(\ref{EQN:OPTIMIZATION_PROBLEM_EXAMPLE}) one sees that for a given problem one must select a FOM, $\Phi(\xi)$, which provides a reliable measure of the performance of the design. Further, one must select a set of functions, $c_i$ and $c_j$, providing reliable measures of all constrains associated with the design problem.\footnote{The function $\Phi$ is sometimes called the performance indicator, the objective function or the objective functional.} \\

\noindent \textit{Considering again our baseline example, a simple and reliable measure of how well a metalens focusses TM-polarized light impinging on it at normal incidence, is obtained by modelling this process for a fixed input power and evaluating the magnitude of the electromagnetic field intensity at the focal point as,}
 
\begin{eqnarray}
\Phi = \vert \textbf{E}(\textbf{r}_{\text{FP}},\varepsilon_r(\textbf{r})) \vert^2 = \vert H_z(\textbf{r}_{\text{FP}},\varepsilon_r(\textbf{r})) \vert^2, \label{EQN:OBJECTIVE_EXAMPLE}
\end{eqnarray}

\noindent \textit{where $\textbf{E}(\textbf{r}_{\text{FP}},\varepsilon_r(\textbf{r}))$ denotes the electric field at the focal point $\textbf{r}_{\text{FP}}$, resulting from a TM-polarized illumination of the material distribution $\varepsilon_r(\textbf{r})$, constituting the metalens and its surrounding environment modelled by solving eq.~(\ref{EQN:EM_HELMHOLTZ_2D}) assuming TM-polarization.} \\
	
 \textcolor{black}{Alternative FOMs for the design problem could be the powerflow through the focal spot, or the integral over the focal plane of the difference between the Airy disc that would be formed by an ideal lens and the field profile formed by the lens under design. All these FOMs can be written as simple functions of the electric and/or magnetic fields evaluated in points, lines or areas.} \\

\noindent \textcolor{black}{When formulating a TopOpt problem, the state equation(s) to be solved (e.g. eq.~(\ref{EQN:EM_HELMHOLTZ_2D})) may be thought of as a set of equality constraints written as,}

\begin{eqnarray} \label{EQN:STATE_EQUATIONS}
\mathcal{L}_k(\textbf{x}_k) = \textbf{f}_k, \ \ k \in \lbrace 1,2,... \mathcal{N}_k\rbrace, \ \ \mathcal{N}_k \in \mathbb{N},
 \end{eqnarray}

\noindent where $\mathcal{L}_k$ is an operator applying the effect of the physical system to the state field $\textbf{x}_k$ for the excitation $\textbf{f}_k$. \\

\noindent \textit{\textcolor{black}{For our baseline example,} having identified our FOM and a way of computing it, \textcolor{black}{by solving eq.~(\ref{EQN:EM_HELMHOLTZ_2D})}, we may now write the design problem as the following optimization problem,}

\begin{eqnarray}
&\underset{\varepsilon_r(\textbf{r})}{\max}& \Phi(H_z(\textbf{r}_{\text{FP}}, \varepsilon_r(\textbf{r}))) \label{EQN:OPTIMIZATION_PROBLEM_GENERAL} \\
&\mathrm{s.t.}& \mathcal{L}_{EM}\left(H_z(\textbf{r}),\varepsilon_r(\textbf{r})\right) = f_z(\textbf{r}). \nonumber
\end{eqnarray}

\noindent To solve optimization problems of the form in eq.~(\ref{EQN:OPTIMIZATION_PROBLEM_EXAMPLE}) using TopOpt, we utilize the continuous design field, $\xi(\textbf{r})$, to interpolate the material parameters in the state equation between the background material(s) and the material(s) constituting the structure under design\footnote{Depending on the problem at hand different material interpolation schemes should be used \cite{CHRISTIANSEN_VESTER_2019}.}. \\

\noindent \textit{In our example, we use a scheme that is linear in $\xi$, to interpolate between air and silicon},

\begin{eqnarray} \label{EQN:LINEAR_INTERPOLATION}
\varepsilon_r(\xi(\textbf{r}))) = \varepsilon_{r,Si} + \xi(\textbf{r}) \left( \varepsilon_{r,Air} - \varepsilon_{r,Si} \right),
\end{eqnarray}

\noindent \textit{where $\varepsilon_{r,Si}$ and $\varepsilon_{r,Air}$ denote the relative permittivity of silicon and air, respectively. That is, we have that $\xi = 0 \Leftrightarrow \varepsilon_r = \varepsilon_{r,Si}$ and $\xi = 1 \Leftrightarrow \varepsilon_r = \varepsilon_{r,Air}$.}  \\

\noindent The introduction of the interpolation function adds an equality constraint to the optimization problem, as well as the following two inequality constraints, bounding $\xi$,

\begin{eqnarray}
0 \leq \xi(\textbf{r}) \leq 1, \ \ \textbf{r} \in \Omega.
\end{eqnarray}

\noindent \textit{Including the material interpolation, the final optimization problem representing our baseline metalens design problem example may now be written as}\footnote{Note that both equality constraints in eq.~(\ref{EQN:OPTIMIZATION_PROBLEM_ITERATION_0}) are satisfied implicitly (within the accuracy of the numerical model used to solve/evaluate them) as $H_z(\textbf{r})$ is computed by solving the first constraint for a given $\varepsilon_r(\textbf{r})$, which in turn is computed by inserting $\xi(\textbf{r})$ in the second constraint.},

\begin{eqnarray}
\underset{\xi(\textbf{r})}{\max} &\Phi(H_z(\textbf{r}_{\text{FP}}, \varepsilon_r(\xi(\textbf{r}))))&  \nonumber \\
\mathrm{s.t.} &\mathcal{L}_{EM}\left(H_z(\textbf{r}),\varepsilon_r(\xi(\textbf{r}))\right) = f_z(\textbf{r})& \label{EQN:OPTIMIZATION_PROBLEM_ITERATION_0} \\
&\varepsilon_r(\xi(\textbf{r}))) = \varepsilon_{r,Si} + \xi(\textbf{r}) \left( \varepsilon_{r,Air} - \varepsilon_{r,Si} \right).& \nonumber \\
&0 \leq \xi(\textbf{r}) \leq 1.& \nonumber
\end{eqnarray}

\noindent In order to solve the continuous constrained optimization problem efficiently TopOpt utilizes gradient-based algorithms\footnote{The Method of Moving Asymptotes \cite{SVANBERG_2002} is often used, as it is efficient for problems with a large design space and few constraints.}. As the name suggests, such algorithms require knowledge of the gradients (sensitivities) of $\Phi$, $c_i$ and $c_j$ with respect to the design field $\xi$. These sensitivities may be approximated naively using finite differences, which entails solving the relevant system(s) of equations for perturbations of each of the design variables in turn at each design iteration. However, doing so is in most cases prohibitively time consuming due to the large number of equations that must be solved. Rather than using finite differences it is advisable to use adjoint sensitivity analysis, an approach which only requires solving a single (adjoint) equation for the FOM and one for each constraint in the optimization problem, independent of the size of the design space. Adjoint sensitivity analysis has been used for TopOpt in the context of mechanical engineering for decades, since the earliest works \cite{BENDSOE_KIKUCHI_1988,TORTORELLI_ET_AL_1994}, and in photonics engineering as detailed in the 2011 review by \cite[Jensen and Sigmund]{JENSEN_SIGMUND_2011}. As such, adjoint sensitivity analysis serves as one of the cornerstones of Topology Optimization. Detailed derivations of adjoint sensitivity analysis in the context of photonics may be found in \cite[Keraly et al]{KERALY_2013} and in \cite[Niederberger et al]{NIEDERBERGER_ET_AL_2014}. 

\subsection{A note on adjoint sensitivity analysis} \label{SEC:ADJOINT}

Adjoint sensitivity analysis may be performed for the analytical equations describing the model problem before numerical discretization or for the discretized system used in the numerical model. The former has the advantage that it is often simple to derive and implement the equations, while the latter has the advantage that the gradients obtained from solving the adjoint problems are exact in terms of the discretized model.

\subsection{A note on optimality} \label{SEC:OPTIMALITY}

\textcolor{black}{We find it appropriate to stress that no mathematical optimization method, be it \emph{global-optimization} based like a genetic algorithm \cite{GOLDBERG_1989}, \emph{artificial-intelligence} based  \cite{RUSSELL_2010} or \emph{gradient} based \cite{SVANBERG_1987}}, is able to guarantee global optimality of the solution to a non-convex optimization problem. Hence, there is no guarantee that the final design field constitutes the best possible solution to the design problem at hand, unless all possible design permutations have been tested. At most, a mathematical optimization method is able to guarantee local optimality, by for example ensuring that the solution fulfils the KKT \textcolor{black}{(Karush-Kuhn-Tucker)} conditions \cite{NOCEDAL_2006}. Therefore the authors caution the reader against the use of the word \emph{optimal} when describing a structure designed using any optimization-based design method. When the performance of the structure is discussed, the authors propose using the word \emph{optimized}, as well as providing a measure of the structures performance relative to theoretical limits or relative to references from the literature, \textcolor{black}{c.f. \cite{MOLESKY_ET_AL_2020,GUSTAFSSON_ET_AL_2020,MICHON_ET_AL_2019}}. 

\subsection{A note on the design uniqueness} \label{SEC:GEOMETRY}

For a number of photonic design problems the authors have found that, depending on the initial guess for the design field, TopOpt is able to identify several qualitatively different designs that perform similarly in terms of the FOM. Hence, the final design geometry may depend strongly on the initial guess without the value of the FOM changing significantly. 

\subsection{A note on parameter tuning} \label{SEC:PARAMETER_TUNING}

All optimization-based inverse design methods use a set of parameters, which must be tuned for each design problem. As a result, one cannot expect to get quality results by simply applying a method "out of the box", without a-priori knowledge and experience with the physics problem at hand as well as with the inverse design method itself.

\section{Software}
 
To help the reader start using Topology Optimization for photonic applications, as well as reproduce the results presented in this paper, a set of COMSOL Multiphysics models (v 5.5) are made available along with this article.\footnote{Note that a COMSOL Multiphysics licence is needed to use the software.} Executing the studies in these models without any modifications will reproduce the data used to create Figs.~\ref{FIG:METALENS_CASE1_ITERATION1}-\ref{FIG:BEAM_SPLITTER_RESULTS}. Readers are invited to use these models as the starting point for their own applications. A brief description of the model \verb!MetalensCase1.mph! may be found in Appendix \ref{APN:COMSOL_MODEL_DESCRIPTION}.

\section{Model Problem: The Metalens}

\noindent As the first design example, we consider a focusing problem. In particular we show how to apply TopOpt to design monochromatic and polychromatic cylindrical metalenses capable of focusing Gaussian-enveloped, TM-polarized plane waves at normal incident into a point. The metalenses consist of a region of silicon and air, placed on top of a massive block of silicon in an air background. We start by considering monochromatic focusing (Sec.~\ref{SEC:METALENS_CASE_1}) and demonstrate the beneficial effects of introducing artificial attenuation (Sec.~\ref{SEC:METALENS_CASE_2}) as well as filtering and thresholding operations (Sec.~\ref{SEC:METALENS_CASE_3}) in the design procedure. We then expand the formulation to consider the case of polychromatic focusing (Sec.~\ref{SEC:METALENS_CASE_4}), hereby showing how simple it can be to design broadband cylindrical metalenses using TopOpt. 

\begin{figure}[h!]
	\centering
	{
		\includegraphics[width=0.8\textwidth]{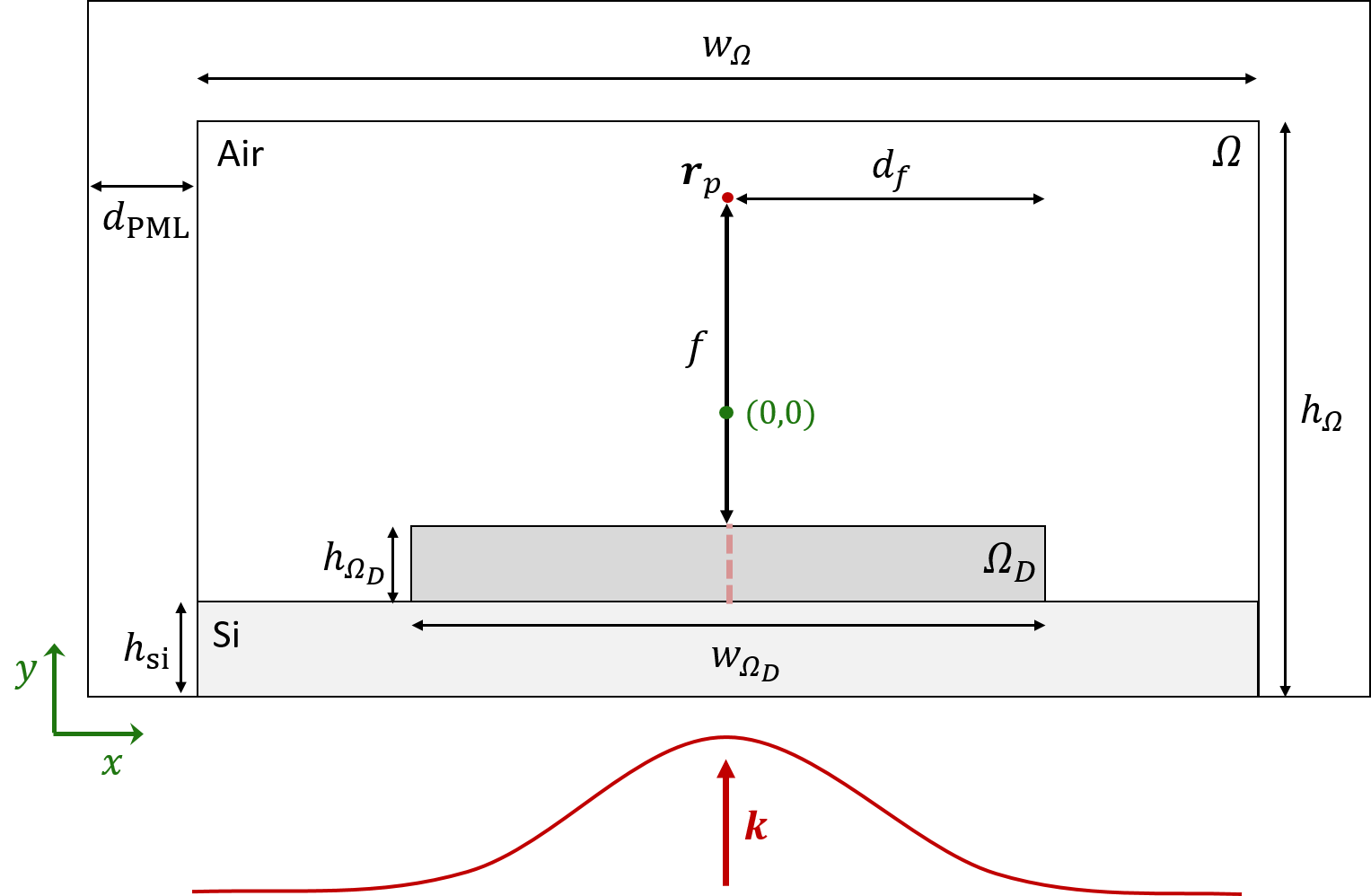} \caption{The model domain, $\Omega$, used for the examples in sections~\ref{SEC:METALENS_CASE_1}-\ref{SEC:METALENS_CASE_4}. Values for all lengths defined in the sketch are found in Tab.~\ref{TAB:METALENS_GEOMETRY}. \label{FIG:META_LENS_MODEL_PROBLEM}}
	}
\end{figure}

The model problem is sketched in Fig.~\ref{FIG:META_LENS_MODEL_PROBLEM} and consists of the model domain $\Omega$ of height, $h_{\Omega}$, and width, $w_{\Omega}$. The domain is thresholded by a perfectly matched layer (PML) \cite{BERENGER_1994} of depth $d_{\mathrm{PML}}$ on three of four sides, with first order scattering boundary conditions imposed on the outside of the PML region. On the lower boundary a TM-polarized planewave, localized using a Gaussian envelope, is introduced using a first-order scattering boundary condition. The choices of boundary conditions made here are made for simplicity and more elaborate boundary conditions may be applied for a number of reasons, such as to obtain the most accurately modelling of the operating conditions of the lens. The incident TM-polarized field has its $H_z$ component on the boundary given as, 

\begin{eqnarray}
H_z = \frac{i}{\mu_0 \omega} \cdot \nabla_y E_x, \ \ E_x = \exp(-x^2/\Delta x_g^2)\exp(-i\textbf{k}\cdot\textbf{r}), \ \ \textbf{k} = \frac{2 \pi}{\lambda} \langle 0, 1 \rangle
\end{eqnarray}

A designable region $\Omega_D$ of height, $h_{\Omega_D}$, and width, $w_{\Omega_D}$, is placed on top of a silicon slab of height $h_{\mathrm{Si}}$ and width $w_{\mathrm{Si}} = w_{\Omega}$. The material distribution in $\Omega_D$ is sought tailored to focus the TM-polarized planewave impinging on the region from the silicon into the focal point, $\textbf{r}_p$, situated a distance $f$ (the focal distance) above the top of $\Omega_D$ and a distance $d_f$ from the right boundary of $\Omega_D$. By selecting $f$ and $w_{\Omega_D}$ one determines the numerical aperture of the lens, NA, as,

\begin{eqnarray}
\mathrm{NA} = \sin\left(\arctan\left(\frac{w_{\Omega_D}}{2f}\right)\right)
\end{eqnarray}

\begin{table}[h!]
\centering{
	\begin{tabular}{l|cccccccc}
		{\ul } & $w_{\Omega}$ [nm] & $h_{\Omega}$ [nm] & $d_{\mathrm{PML}}$ [nm] & $w_{\Omega_D}$ [nm] & $h_{\Omega_D}$ [nm] & $h_{\mathrm{Si}}$ [nm] & $d_f$ [nm] & $f$ [nm] \\ \hline
			\\ [-1em]
		Case 1-4 & 4800  & 1816.2  & 550 & 3000 & 250 & 125 & 1500 & 726.5 \\
	\end{tabular}
\caption{Values for quantities in Fig. \ref{FIG:META_LENS_MODEL_PROBLEM} for the cases considered in the following.} \label{TAB:METALENS_GEOMETRY}
}
\end{table}

\begin{table}[h!]
	\centering{
		\begin{tabular}{l|cccccccc}
			{\ul } & $\lambda$ [nm] & $n_{Si}$ & $n_{BG}$ & $\Delta x_g$ [nm] & NA & $h_e$ [nm] & $h_{\Omega_D}$ [nm] \\ \hline
			\\ [-1em]
			Case 1-4 & 550  & 3.48  & 1.00   & 1500 & 0.9 &  $\lambda / (10 n(\textbf{r})$) & 5   \\
		\end{tabular}
		\caption{Values for the physical, material and discretization parameters for the cases considered in the following. The wavelength $\lambda$, the refractive indices for Si and air, $n_{Si}$ and $n_{BG}$, the element size used to discretize the model away from the design $h_e$ and in the design domain $h_{dx}$,  $h_{dy}$. The width of the Gaussian-envelope $\Delta x_g$ and the numerical aperture NA.} \label{TAB:METALENS_OTHER_PARAMETERS}
	}
\end{table}

\begin{table}[h!]
	\centering{
		\begin{tabular}{l|cccccc}
			{\ul } & $\xi_{ini}$  & $\alpha_{a}$ & $\beta_{ini}$ & $\eta$ & $r_f$ & $n_{\mathrm{iter}}$ \\ \hline
			\\ [-1em]
			Case 1 & 0.5 & N/A & N/A & N/A & N/A & 1000 \\
			\hline
			\\ [-1em]
			Case 2 & 0.5 & 1.0 & N/A & N/A & N/A & 1000 \\
			\hline
			\\ [-1em]
			Case 3-4 & 0.5 & 1.0 & 1.0 & 0.5 & 80 [nm] & 1000 \\
		\end{tabular}
		\caption{Values for quantities related to manipulating the design field and solving the optimization problems. The initial value for the design field $\xi_{ini}$, the (initial) filter strength $\beta_{ini}$ and threshold level $\eta$, the filter radius $\textbf{r}_f$ and the fixed number of inner iterations taken to solve the optimization problem, $n_{\mathrm{iter}}$.} \label{TAB:METALENS_OPTIMIZATION_PARAMETERS}
	}
\end{table}

Table \ref{TAB:METALENS_GEOMETRY} lists the values of the quantities defined in Fig. \ref{FIG:META_LENS_MODEL_PROBLEM} for the cases considered in the following. The physical, material and discretization parameters used in the cases are listed in Tab.~\ref{TAB:METALENS_OTHER_PARAMETERS}. Finally, Tab.~\ref{TAB:METALENS_OPTIMIZATION_PARAMETERS} lists parameters related to manipulating the design field and solving the optimization problem.

\subsection{Case 1: Naive approach} \label{SEC:METALENS_CASE_1} 

\noindent A first naive approach to designing a metalens using TopOpt is to solve eq.~(\ref{EQN:OPTIMIZATION_PROBLEM_ITERATION_1}), derived in Sec.~\ref{SEC:OPTIMIZATION_PROBLEM}, using the parameter choices listed in Tabs.~\ref{TAB:METALENS_GEOMETRY}-\ref{TAB:METALENS_OPTIMIZATION_PARAMETERS}. 
\begin{eqnarray}
\underset{\xi(\textbf{r})}{\max} &\Phi(H_z(\textbf{r}, \varepsilon_r(\xi(\textbf{r}))))&  \nonumber \\
\mathrm{s.t.} &\mathcal{L}_{EM}\left(H_z(\textbf{r}),\varepsilon_r(\xi(\textbf{r}))\right) = f_z(\textbf{r})& \label{EQN:OPTIMIZATION_PROBLEM_ITERATION_1} \\
&\varepsilon_r(\xi(\textbf{r}))) = \varepsilon_{r,Si} + \xi(\textbf{r}) \left( \varepsilon_{r,Air} - \varepsilon_{r,Si} \right).& \nonumber \\
&0 \leq \xi(\textbf{r}) \leq 1.& \nonumber
\end{eqnarray}

\begin{figure}[h!]
	\centering
	{
		\includegraphics[width=1.0\textwidth]{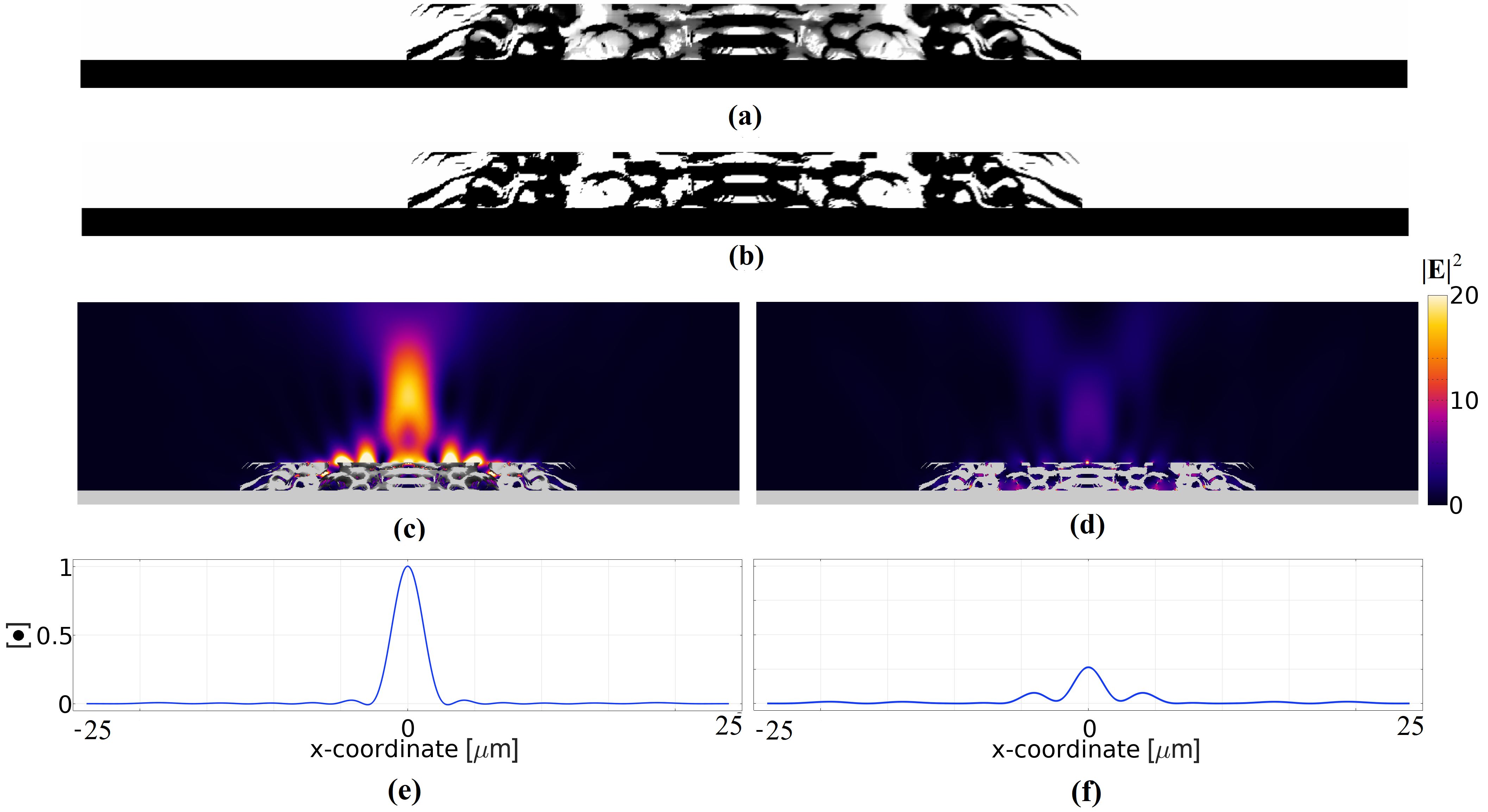} \caption{\textbf{(a)} Design field, $\xi(\textbf{r})$, obtained by solving the optimization problem in eq.~(\ref{EQN:OPTIMIZATION_PROBLEM_ITERATION_1}), black(white) corresponds to $\xi = 1(\xi = 0)$. \textbf{(b)} Thresholded design field, $\xi(\textbf{r})>0.5$. \textbf{(c-d)} The electric field intensity $\vert \textbf{E} \vert^2$ (thermal) with the metalens overlaid (gray) for the \textbf{(c)} optimized design in \textbf{(a)} and \textbf{(d)} the thresholded design in \textbf{(b)}. \textbf{(e-f)} The max-normalized power flow, $\langle \textbf{P} \rangle$,  through the focal plane for the \textbf{(e)} optimized design in \textbf{(a)} and \textbf{(f)} the thresholded design in \textbf{(b)}. \label{FIG:METALENS_CASE1_ITERATION1}}
	}
\end{figure}

The reader may solve this problem by executing the study named \textit{Optimization} in the COMSOL model \verb!MetalensCase1.mph!. The final design field, $\xi(\textbf{r})$, obtained by solving the optimization problem, is shown in Fig.~\ref{FIG:METALENS_CASE1_ITERATION1}(a), with black corresponding to silicon and white corresponding to air. 

An immediate observation, is that $\xi(\textbf{r})$ contains several gray regions of intermediate values, i.e. $\xi(\textbf{r}) \in ]0,1[$. These regions appear during the design process, because they allow the metalens to manipulate the wavelength and phase of the electromagnetic field locally with high precision, thus enabling enhancement of the focussing efficiency. While these regions \textcolor{black}{may be} beneficial to the performance of the design, they are non-physical, or at least impractical to realize. In order to fabricate a metalens from the design one would have to post-process it by removing all intermediate values of  $\xi(\textbf{r})$, or do deep sub-wavelength perforations of the silicon to approximate the refractive index in the gray regions, to obtain a design consisting solely of silicon (black) and air (white). 

A simple post-processing approach, to obtain a physically-realizable design, is to threshold $\xi(\textbf{r})$ at 0.5. Doing so results in the crisp black and white design presented in Fig.~\ref{FIG:METALENS_CASE1_ITERATION1}(b). However, the threshold operation results in a design, which is significantly different from the optimized design. Hence, it is unlikely that the two designs will perform equally well. In fact the FOM drops from $\Phi \approx 18.2 \ [\mathrm{V}^2/\mathrm{m}^2]$ for the design in Fig.~\ref{FIG:METALENS_CASE1_ITERATION1}(a) to $\Phi \approx 4.7 \ [\mathrm{V}^2/\mathrm{m}^2]$ for the thresholded design in Fig.~\ref{FIG:METALENS_CASE1_ITERATION1}(b). From a visual investigation of the $|E|$-field emitted from the original design, seen in Fig.~\ref{FIG:METALENS_CASE1_ITERATION1}(c), and the thresholded design, seen in Fig.~\ref{FIG:METALENS_CASE1_ITERATION1}(d), it is clear that the E-fields are different and that significantly less energy is transmitted through the thresholded design. Considering the max-normalized, time-averaged powerflow, $\langle \textbf{P} \rangle$, normal to the focal plane in Fig.~\ref{FIG:METALENS_CASE1_ITERATION1}(e)~and~\ref{FIG:METALENS_CASE1_ITERATION1}(f) (a metric often used to determine lens performance), it is clearly seen that the power focused at the focal point drops significantly when thresholding the design.

From this analysis it is clear that the first, naive, approach to applying TopOpt, that does not include any regularization of the design field, risks yielding poor results because intermediate values of $\xi(\textbf{r})$ are allowed in the design process but not in the final physical metalens. In the next iteration we demonstrate a modification ensuring a final optimized design without intermediate values, namely a scheme for implicitly penalizing intermediate values of $\xi$.

\subsection{Case 2: Artificial attenuation} \label{SEC:METALENS_CASE_2}

As we only care about maximizing $\vert \textbf{E} \vert^2$ at the focal spot, a simple yet effective approach to eliminate intermediate design values, is to introduce artificial attenuation of the electromagnetic field for any intermediate value of $\xi$ (also called penalization damping or pamping \cite{JENSEN_SIGMUND_2005}) by adding an imaginary part to the interpolation scheme for the relative permittivity as,

\begin{eqnarray}
\varepsilon_r(\xi) = 1 + \xi \left( \varepsilon_r - 1 \right) - \mathrm{i} \alpha_{a} \xi (1-\xi).
\end{eqnarray}

This way of physical penalization of intermediate design field values is much preferred to explicit penalization, like adding $\int_{\Omega_D} \xi (1-\xi) \mathrm{d}\textbf{r}$ to the FOM, which tends to get the result stuck in bad local minima. \textcolor{black}{The} added artificial attenuation means that if $\xi$ takes an intermediate value anywhere in $\Omega_D$ where there is a non-zero electric field, it will result in less energy propagating through the metalens, which is ultimately detrimental to maximizing $\Phi$. The modified optimization problem in eq.~(\ref{EQN:OPTIMIZATION_PROBLEM_ITERATION_2}) may be solved by the reader by executing the study named \textit{Optimization} in the model \verb!MetalensCase2.mph!.

\begin{eqnarray}
\underset{\xi(\textbf{r})}{\max} &\Phi(H_z(\textbf{r}), \varepsilon_r(\xi(\textbf{r})))&  \nonumber \\
\mathrm{s.t.} &\mathcal{L}_{EM}\left(\varepsilon_r(\xi(\textbf{r})),H_z(\textbf{r})\right) = \textbf{f}_z(\textbf{r})& \label{EQN:OPTIMIZATION_PROBLEM_ITERATION_2} \\
&\varepsilon_r(\xi) = 1 + \xi \left( \varepsilon_r - 1 \right) - \mathrm{i} \alpha_{a} \xi (1-\xi).& \nonumber \\
&0 \leq \xi(\textbf{r}) \leq 1.& \nonumber
\end{eqnarray}

\begin{figure}[h!]
	\centering
	{
		\includegraphics[width=1.0\textwidth]{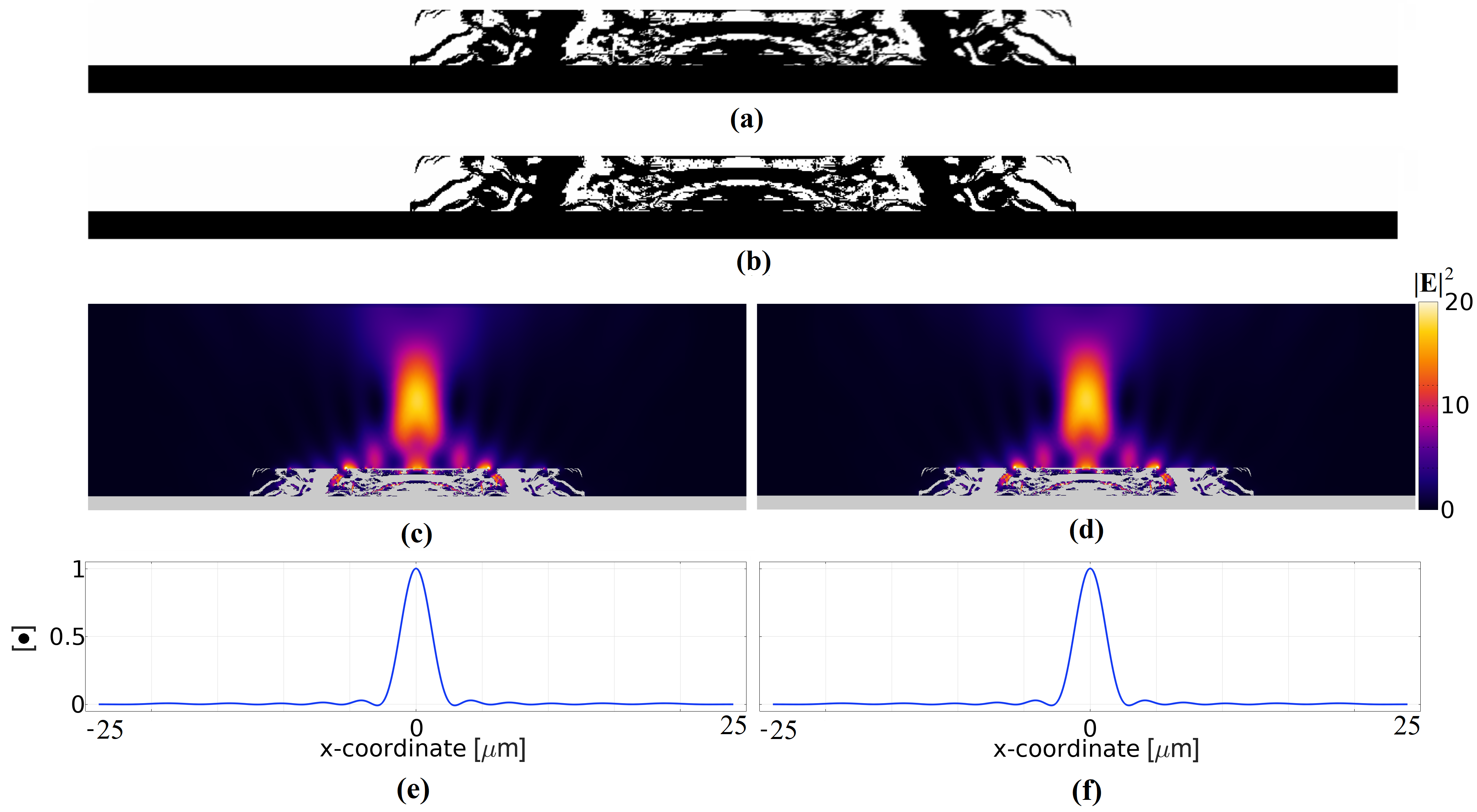} \caption{\textbf{(a)} Design field, $\xi(\textbf{r})$, obtained by solving the optimization problem in eq.~(\ref{EQN:OPTIMIZATION_PROBLEM_ITERATION_2}), black(white) corresponds to $\xi = 1(\xi = 0)$. \textbf{(b)} Thresholded design field, $\xi(\textbf{r})>0.5$. \textbf{(c-d)} The electric field intensity $\vert \textbf{E} \vert^2$ (thermal) with the metalens overlaid (gray) for the \textbf{(c)} optimized design in \textbf{(a)} and \textbf{(d)} the thresholded design in \textbf{(b)}. \textbf{(e-f)} The max-normalized power flow through the focal plane for the \textbf{(e)} optimized design in \textbf{(a)} and \textbf{(f)} the thresholded design in \textbf{(b)}. \label{FIG:METALENS_CASE1_ITERATION2}}
	}
\end{figure}

Executing this study yields the $\xi(\textbf{r})$-field shown in Fig.~\ref{FIG:METALENS_CASE1_ITERATION2}(a). From the figure it is observed that $\xi(\textbf{r})$ now contains (almost) no intermediate values. In fact, thresholding $\xi(\textbf{r})$ at 0.5 results in the (almost) identical blank and white design, shown in Fig.~\ref{FIG:METALENS_CASE1_ITERATION2}(b). 

Evaluating the performance of the two designs in Fig.~\ref{FIG:METALENS_CASE1_ITERATION2}\textbf{(a-b)}, one obtains near identical values of the FOM, namely $\Phi \approx 17.84$. Inspecting the $\vert \textbf{E} \vert^2$-fields for the optimized and thresholded design in Fig.~\ref{FIG:METALENS_CASE1_ITERATION2}(c-d), they indeed look (near) identical. Considering the time averaged powerflow through the focal plane for the two designs, shown in Fig.~\ref{FIG:METALENS_CASE1_ITERATION2}(e-f), it is seen that the power focused at the focal point does not drop when thresholding the design.  

Thus, from a performance perspective Case 2 yields an optimized monochromatic cylindrical metalens, which is physically realizable as it consists solely of silicon and air. However, if we look at the design in Fig.~\ref{FIG:METALENS_CASE1_ITERATION2}(b), we observe multiple pixel-by-pixel varying material regions as well as several tiny and narrow features, which raises the following problems. First, the quality of the numerical modelling. Having pixel-by-pixel varying material parameters is in general not good numerical modelling and may, in some cases, result in poor accuracy of the numerical model. Second, it is unlikely that it is possible to accurately manufacture designs with extremely small and rapidly varying features. Third, a design with such features is likely not mechanically stable.

\textcolor{black}{Multiple} methods for amending the problem of tiny and single-pixel features have been developed in the context of mechanical engineering \cite{WANG_ET_AL_2011,LAZAROV_ET_AL_2016}. In the next iteration we demonstrate a conceptually simple approach.

\subsection{Case 3: Filter and threshold} \label{SEC:METALENS_CASE_3}

This case introduces a well known filtering and thresholding procedure \cite{GUEST_ET_AL_2004} to control spatial design-field variations. This is a simple, yet effective, way of introducing a weak sense of lengthscale into the design and remedies poor numerical modelling with single-pixel features. The smoothing filter is applied by solving the following auxiliary Partial Differential Equation (PDE) in the design domain $\Omega_D$ for the filtered design field, $\tilde{\xi}$, with the original design field, $\xi$, as input and homogeneous Neumann boundary conditions on all \textcolor{black}{other} boundaries\footnote{Note that the filtering procedure is not extended beyond the design domain for simplicity.} \cite{LAZAROV_2011},

\begin{eqnarray}
-\left(\frac{r_f}{2\sqrt{3}}\right)^2 \nabla \tilde{\xi}(\textbf{r})+\tilde{\xi}(\textbf{r})=\xi(\textbf{r}), \ \ r_f  \geq 0, \ \ \textbf{r} \in \Omega_D. \label{EQN:FILTER_PDE}
\end{eqnarray}

Here $r_f$ denotes the desired spatial filtering radius. By varying the filter radius, $r_f$, it is possible to exert control on the size of the features appearing in the filtered design field. \\

\noindent \textcolor{black}{Next, t}he filtered field is thresholded using a smoothed approximation of the Heaviside function \cite{WANG_ET_AL_2011} to recover a nearly discrete design, 

\begin{eqnarray}
\bar{\tilde{\xi}} = \frac{\tanh(\beta \cdot \eta) + \tanh(\beta \cdot (\tilde{\xi} - \eta))}{\tanh(\beta \cdot \eta) + \tanh(\beta \cdot (1 - \eta))}, \ \ \beta \in [1,\infty[, \ \ \eta \in [0,1].  \label{EQN:THRESHOLD_FUNCTION}
\end{eqnarray}

Here $\beta$ controls the threshold sharpness and $\eta$ controls the threshold value. At $\beta = 1$ the thresholding has little effect, i.e. $\bar{\tilde{\xi}} \approx \tilde{\xi}$. whereas for $\beta$ approaching infinity the thresholded field only takes values of 0 or 1, i.e. $\lim\limits_{\beta \rightarrow \infty}(\bar{\tilde{\xi}}) \in \lbrace 0, 1 \rbrace$. 

The filter and threshold procedure is applied during the solution of the optimization problem using a continuation scheme, where $\beta$ is increased from a relatively low starting value to a relatively high stopping value. The low starting value of $\beta$ allows the design to develop as if no thresholding was performed in the initial stage of solving the optimization problem, while the high stopping value allows a pure black and white design in the later stage of solving the optimization problem. \\

The optimization problem being solved in this case is written as,

\begin{eqnarray}
\underset{\xi(\textbf{r})}{\max} &\Phi(H_z(\textbf{r}), \varepsilon_r(\bar{\tilde{\xi}}(\textbf{r})))&  \nonumber \\
\mathrm{s.t.} &\mathcal{L}_{EM}\left(\varepsilon_r(\bar{\tilde{\xi}}(\textbf{r})),H_z\right)= \textbf{f},& \nonumber \\
&\varepsilon_r(\bar{\tilde{\xi}}) = 1 + \bar{\tilde{\xi}} \left( \varepsilon_r - 1 \right), &  \label{EQN:OPTIMIZATION_PROBLEM_ITERATION_3} \\
&\bar{\tilde{\xi}} = \frac{\tanh(\beta \cdot \eta) + \tanh(\beta \cdot (\tilde{\xi} - \eta))}{\tanh(\beta \cdot \eta) + \tanh(\beta \cdot (1 - \eta))}, & \nonumber \\
&-\left(\frac{r_f}{2\sqrt{3}}\right)^2 \nabla \tilde{\xi}(\textbf{r})+\tilde{\xi}(\textbf{r})=\xi(\textbf{r}),& \nonumber \\
&0 \leq \xi(\textbf{r}) \leq 1.& \nonumber
\end{eqnarray}

\begin{figure}[h!]
	\centering
	{
		\includegraphics[width=1.0\textwidth]{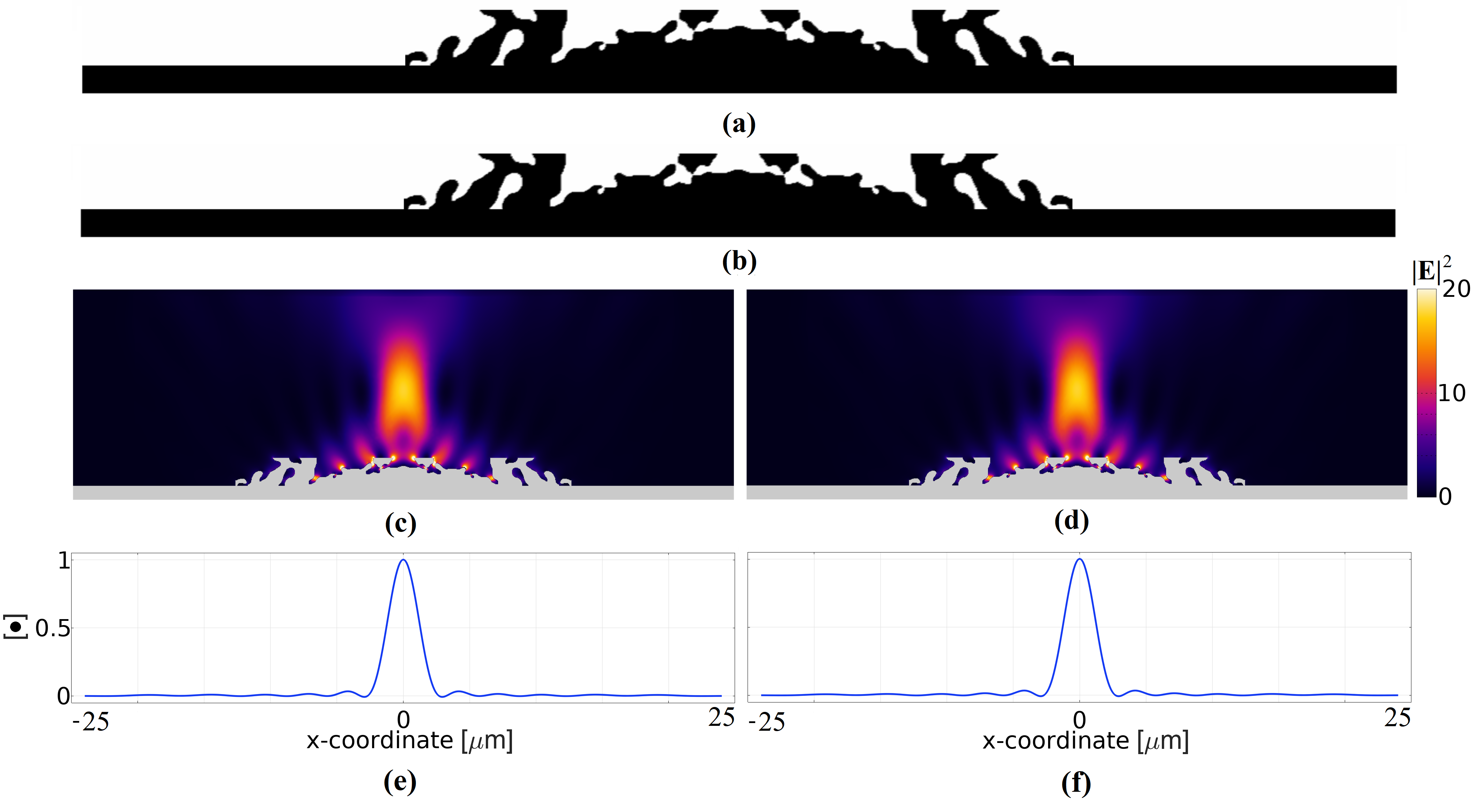} \caption{\textbf{(a)} Design field, $\xi(\textbf{r})$, obtained by solving the optimization problem in eq.~(\ref{EQN:OPTIMIZATION_PROBLEM_ITERATION_3}), black(white) corresponds to $\xi = 1(\xi = 0)$. \textbf{(b)} Thresholded design field, $\xi(\textbf{r})>0.5$. \textbf{(c-d)} The electric field intensity $\vert \textbf{E} \vert^2$ (thermal) with the metalens overlaid (gray) for the \textbf{(c)} optimized design in \textbf{(a)} and \textbf{(d)} the thresholded design in \textbf{(b)}. \textbf{(e-f)} The max-normalized power flow through the focal plane for the \textbf{(e)} optimized design in \textbf{(a)} and \textbf{(f)} the thresholded design in \textbf{(b)}. \label{FIG:METALENS_CASE1_ITERATION3}}
	}
\end{figure}

The reader may solve the problem in eq.~(\ref{EQN:OPTIMIZATION_PROBLEM_ITERATION_3}), using continuation of the $\beta$-values, by executing the study named \textit{Continuation} in the model \verb!MetalensCase3.mph!. Hereby the field, ${\bar{\tilde{\xi}}}(\textbf{r})$, shown in Fig.~\ref{FIG:METALENS_CASE1_ITERATION3}(a), is obtained. It is observed that the final design consists (almost) solely of silicon and air and that this design contains no single-pixel features. Comparing Fig.~\ref{FIG:METALENS_CASE1_ITERATION2}(a) to Fig.~\ref{FIG:METALENS_CASE1_ITERATION3}(a), it is seen that the design consists of significantly fewer and larger features. Thresholding the final design at 0.5 results in the (almost) identical design shown in Fig.~\ref{FIG:METALENS_CASE1_ITERATION3}(b). 

Evaluating the performance of the designs in Fig.~\ref{FIG:METALENS_CASE1_ITERATION3}(a) and Fig.~\ref{FIG:METALENS_CASE1_ITERATION3}(b), one finds that near identical values of the FOM are obtained, namely $\Phi \approx 17.64$. When comparing the value of $\Phi$ with the value for the design from Case 2 in Fig.~\ref{FIG:METALENS_CASE1_ITERATION2}(a), where no filtering was imposed, a decrease of merely $\approx 1\%$ is observed. That is, the removal of the pixel-by-pixel variations had little impact on the performance of the design, whereas the design geometry has been simplified significantly.  

Looking at the $\vert \textbf{E} \vert^2$-field for the original design and the thresholded design, see  Fig.~\ref{FIG:METALENS_CASE1_ITERATION3}{(c-d)}, they look (near) identical. Looking at the differences between the time averaged powerflow through the focal plane in Fig.~\ref{FIG:METALENS_CASE1_ITERATION3}{(e-f)}, it is clearly seen that the power focused at the focal point is (near) identical for the two designs.  

In conclusion, Case 3 presents an optimized monochromatic cylindrical metalens, which consists solely of silicon and air with significantly larger features than those seen for Case 2, which makes fabrication simpler. Further, from a numerical modelling point of view, the design in Case 3 does not contain rapid pixel-by-pixel material variations, which may jeopardize numerical convergence and precision.

\subsection{Case 4: Multiple state-equation optimization} \label{SEC:METALENS_CASE_4}

There exist many applications where a structure is sought designed to maximize a set of figures of merit simultaneously. A simple way of doing this is to agglomerate these into a single FOM e.g. using a p-norm. Another way is to use a min/max formulation \cite{SIGMUND_JENSEN_2003}. In this example we demonstrate the former. 

We consider the problem of designing a metalens for broadband operation. That is, instead of only maximizing the focusing efficiency at a single wavelength, we target three wavelengths in a 100 nm band simultaneously, i.e. $\lambda \in \lbrace 500 \ \mathrm{nm}, 550 \ \mathrm{nm}, 600 \ \mathrm{nm} \rbrace$. To do this we reformulate the FOM as follows,

\begin{eqnarray} \label{EQN:FOM_CASE4}
\tilde{\Phi} = \left( \sum_{i=1}^{\mathcal{N}_\lambda=3} \sqrt{\Phi_i(H_z(\lambda_i,\textbf{r}, \varepsilon_r(\bar{\tilde{\xi}}(\textbf{r}))))} \right)^2,
\end{eqnarray}

\noindent where $\Phi_i$ is defined in eq.~(\ref{EQN:OBJECTIVE_EXAMPLE}). This formulation inherently puts the greatest weight on the $\Phi_i$, which takes the lowest value. In order to evaluate eq.~(\ref{EQN:FOM_CASE4}), a set of $i$ state equations must be solved. Note that the solution of the state equations can in principle be computed in parallel and hence does not need to increase wall clock time.  

Besides the above change to the FOM, the optimization problem being solved remains the same as in eq.~(\ref{EQN:OPTIMIZATION_PROBLEM_ITERATION_3}). The reader may solve the new problem by executing the study named \textit{Continuation} in the model \verb!MetalensCase4.mph!. 

\begin{figure}[h!]
	\centering
	{
		\includegraphics[width=1\textwidth]{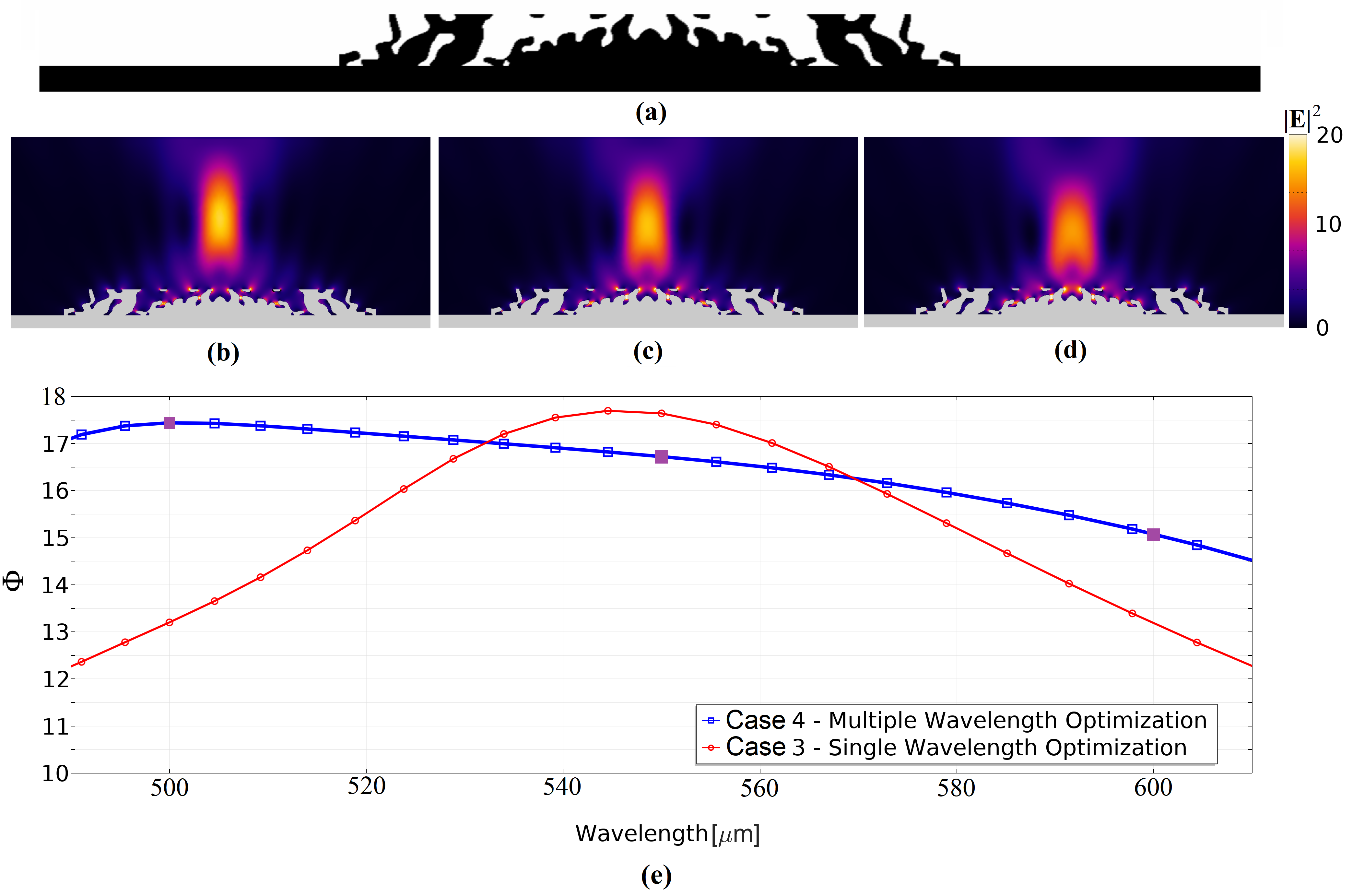} \caption{\textbf{(a)} Design field, $\xi(\textbf{r})$, obtained by solving the optimization problem in eq.~(\ref{EQN:OPTIMIZATION_PROBLEM_ITERATION_1}), black(white) corresponds to $\xi = 1(\xi = 0)$. \textbf{(b-d)}  The electric field intensity $\vert \textbf{E} \vert^2$ (thermal) with the metalens overlaid (gray) for  the optimized design in \textbf{(a)} for \textbf{(b)} $\lambda = 500$ nm, \textbf{(b)} $\lambda = 550$ nm, \textbf{(b)} $\lambda = 600$ nm. \textbf{(e)} The $\Phi$ as a function of wavelength for the design in Fig.~\ref{FIG:METALENS_CASE1_ITERATION4}\textbf{(a)} (blue line) and the design in Fig.~\ref{FIG:METALENS_CASE1_ITERATION3}\textbf{(a)} (red line). \label{FIG:METALENS_CASE1_ITERATION4}}
	}
\end{figure}

Doing so yields the design field,  ${\bar{\tilde{\xi}}}(\textbf{r})$, shown in Fig.~\ref{FIG:METALENS_CASE1_ITERATION4}(a). The $\vert \textbf{E} \vert^2$-fields for the three targeted wavelengths are shown in Figs.~\ref{FIG:METALENS_CASE1_ITERATION4}(b-d). To demonstrate that we have designed a metalens with a better average peformance over the targeted wavelengths compared to only targeting a single wavelength, we evaluate the performance of the design in Fig.~\ref{FIG:METALENS_CASE1_ITERATION4}(a) and the design optimized for the single wavelength $\lambda = 550$ nm, shown in Fig.~\ref{FIG:METALENS_CASE1_ITERATION3}(a), for wavelengths in the interval from 480 nm to 620 nm. The value of the FOM in eq.~(\ref{EQN:OBJECTIVE_EXAMPLE}) as a function of wavelength is plotted in Fig.~\ref{FIG:METALENS_CASE1_ITERATION4}(e). From this figure, it is observed that the design optimized for a single wavelength (unsurprisingly) performs best at that wavelength. However, when considering the full wavelength interval it is clearly observed that the metalens optimized for three wavelengths performs best when averaged over the three wavelengths and also when averaged over the entire interval. Thus, by accepting a performance drop at the central wavelength it is possible to design a lens with significantly better broad-band performance. \textcolor{black}{More constant performance over the frequency interval may be targeted by raising the p-norm value from 2 to a higher value. However, remark that too high values of p (e.g. $p>10$) make the problem highly non-linear, which possibly causes ill-convergence.}

\newpage 
\section{Model Problem: The Demultiplexer} \label{SEC:BEAMSPLITTER}

As an illustration of the broad applicability of TopOpt within photonics design, we consider the design of an optically small photonic demultiplexer (Device footprint $\approx 2.4 \lambda_1^2$), \textcolor{black}{intended} to direct light from a single input waveguide to two different output waveguides depending on the wavelength of the incident light. The example also demonstrates how TopOpt may be used to create designs exhibiting geometric robustness towards near-uniform variations in the geometry. A type of variations similar to those associated with sample over(under) exposure\textcolor{black}{(or etching)} during various nano-fabrication processes \cite{JANSEN_ET_AL_2013,ZHOU_SMO_2015,ERIKSEN_ET_AL_2018}.

\begin{figure}[h!]
	\centering
	{
		\includegraphics[width=1\textwidth]{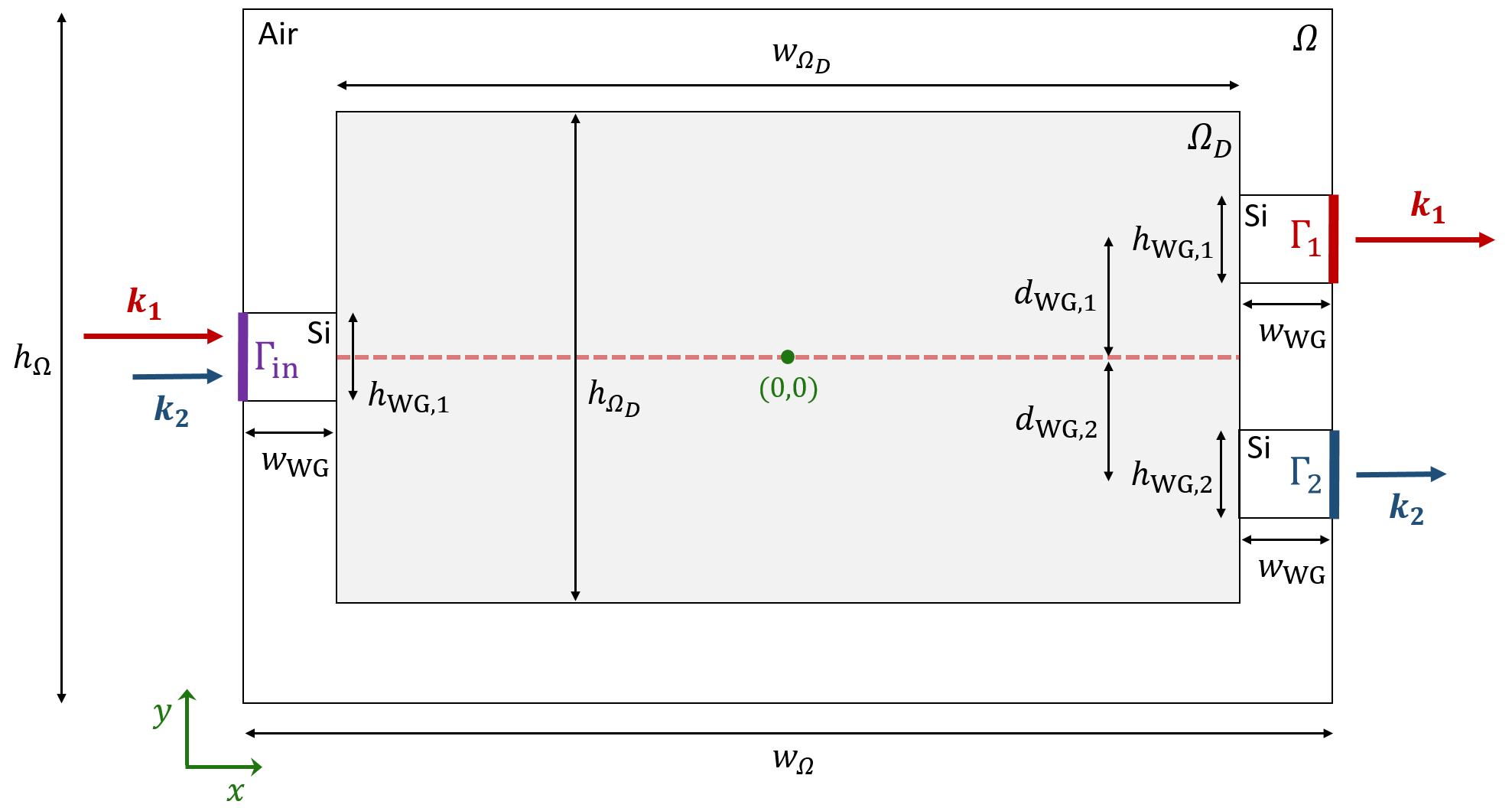} \caption{The model domain $\Omega$ considered in Sec.~\ref{SEC:BEAMSPLITTER}. Values for the lengths are found in Tab.~\ref{TAB:BEAMSPLITTER_GEOMETRY}. \label{FIG:BEAM_SPLITTER_MODEL_PROBLEM}}
	}
\end{figure}

\noindent The model problem considered in this example is shown in Fig.~\ref{FIG:BEAM_SPLITTER_MODEL_PROBLEM} and Tab.~\ref{TAB:BEAMSPLITTER_GEOMETRY} lists values for the length quantities defined on the figure.

\begin{table}[h!]
	\centering{
		\begin{tabular}{l|cccccccccc}
			Parameter & $w_{\Omega}$ & $h_{\Omega}$ & $d_{\mathrm{PML}}$ & $w_{\Omega_D}$ & $h_{\Omega_D}$ & $h_{\mathrm{WG},1}$ & $h_{\mathrm{WG},2}$ & $w_{\mathrm{WG}}$ & $ d_{\mathrm{WG},1}$ & $ d_{\mathrm{WG},2}$ \\ \hline
			\\ [-1em]
			Value [nm] & 6000  & 3000  & 1000 & 2000 & 2000 & 299 & 356 & 2000 & 511 & 483  \\
		\end{tabular}
		\caption{Values for lengths in Fig. \ref{FIG:BEAM_SPLITTER_MODEL_PROBLEM}.} \label{TAB:BEAMSPLITTER_GEOMETRY}
	}
\end{table}

\noindent The model problem setup consists of the model domain $\Omega$ with the height, $h_{\Omega}$, and width, $w_{\Omega}$, contains air as the background medium with the input(output) waveguides and the beam-splitter consisting of silicon. The designable region $\Omega_D$, constituting the demultiplexer, has the height, $h_{\Omega_D}$, and width, $w_{\Omega_D}$, and is placed at the center of $\Omega$. The material distribution in $\Omega_D$ is sought tailored to maximize the time averaged powerflow from the input waveguide into one of the two output waveguides, depending on the wavelength of the light. The model domain is thresholded by a PML of depth $d_{\mathrm{PML}}$ on the left side, with a first order scattering boundary condition imposed on the outside of the PML region. First order scattering boundary conditions are also imposed on the remaining three sides of  $\Omega$. On the left boundary a TE-polarized planewave, localized using a Gaussian envelope, is introduced into the input waveguide of height, $h_{\mathrm{WG,1}}$, and width, $w_{\mathrm{WG}}$. The incident TE-polarized field has the $E_z$ component given as, 

\begin{eqnarray}
E_z = \exp(-y^2/\Delta y_g^2)\exp(-i\textbf{k}_j\cdot\textbf{r}), \ \ \textbf{k}_j = \frac{2 \pi}{\lambda_j} \langle 0, 1 \rangle, \ \ j \in \lbrace 1,2 \rbrace.
\end{eqnarray}

\noindent More advanced boundary conditions may be applied for a number of reasons and the choices of the boundary conditions made here are made purely for simplicity. 

\noindent We demonstrate how to optimize the design to achieve performance robustness towards (near) uniform geometric perturbations by using the double filtering method \cite{CHRISTIANSEN_SMO_2015}. In brief, the method consists of applying the filter and threshold procedure described in Sec.~\ref{SEC:METALENS_CASE_3} twice on the design field, where in the second application three different threshold values are applied to obtain three different realizations of the design fields corresponding to under(over) etching. Six state equations are then solved, two for each of the three realizations of the design fields, corresponding to the two wavelengths targeted by the demultiplexer. The design problem may be formulated as the following constrained optimization problem,

\begin{eqnarray}
\underset{\xi(\textbf{r})}{\max}& \Phi = \sum_{j=1}^{\mathcal{N}_\lambda=2} \sum_{k=1}^{\mathcal{N}_k=3} \sqrt{\left(\int_{\Gamma_j} \langle  \textbf{P}(\lambda_j, \bar{\tilde{\bar{\tilde{\xi}}}}(\textbf{r},\eta_{2,k}))\rangle \mathrm{d}\textbf{r} \right)} \ \ j \in \lbrace 1,2 \rbrace, \ \ k \in \lbrace 1,2,3 \rbrace,&  \nonumber \\
\mathrm{s.t.} &\mathcal{L}_{EM}\left(\lambda_j,\varepsilon_r(\bar{\tilde{\bar{\tilde{\xi}}}}(\textbf{r},\eta_{2,k})),E_z)\right) \textbf{x} = \textbf{f}(\lambda_j),& \nonumber  \\
&\varepsilon_r(\bar{\tilde{\bar{\tilde{\xi}}}}) = 1 + \bar{\tilde{\bar{\tilde{\xi}}}} \left( \varepsilon_r - 1 \right) + \mathrm{i} \bar{\tilde{\bar{\tilde{\xi}}}} (1-\bar{\tilde{\bar{\tilde{\xi}}}}),& \nonumber \\
&\bar{\tilde{\bar{\tilde{\xi}}}} = \frac{\tanh(\beta_2 \cdot \eta_{2,k}) + \tanh(\beta_2 \cdot (\tilde{\bar{\tilde{\xi}}} - \eta_{2,k}))}{\tanh(\beta_2 \cdot \eta_{2,k}) + \tanh(\beta_2 \cdot (1 - \eta_{2,k}))},& \label{EQN:OPTIMIZATION_PROBLEM_BEAM_SPLITTER}  \\
&-\left(\frac{r_{f,2}}{2\sqrt{3}}\right)^2 \nabla \tilde{\bar{\tilde{\xi}}}(\textbf{r})+\tilde{\bar{\tilde{\xi}}}(\textbf{r})=\bar{\tilde{\xi}}(\textbf{r}),& \nonumber \\
&\bar{\tilde{\xi}} = \frac{\tanh(\beta_1 \cdot \eta_1) + \tanh(\beta_1 \cdot (\tilde{\xi} - \eta_1))}{\tanh(\beta_1 \cdot \eta_1) + \tanh(\beta_1 \cdot (1 - \eta_1))},& \nonumber \\
&-\left(\frac{r_{f,1}}{2\sqrt{3}}\right)^2 \nabla \tilde{\xi}(\textbf{r})+\tilde{\xi}(\textbf{r})=\xi(\textbf{r}),& \nonumber \\
&0 \leq \xi(\textbf{r}) \leq 1.& \nonumber
\end{eqnarray}

The physical, material and discretization parameters used in the model are listed in Tab.~\ref{TAB:BEAMSPLITTER_OTHER}.

\begin{table}[h!]
	\centering{
		\begin{tabular}{l|ccccccccc}
			Parameter & $\lambda_1$ & $\lambda_2$ & $n_{\textbf{S}i}$ & $n_{BG}$ & $h_e$ & $h_{dx}$ ($h_{\Omega_D}$) & $h_{dy}$ ($h_{\Omega_D}$)  & $\Delta y_g$ \\ \hline
			\\ [-1em]
			 Value & 1300 nm  & 1550 nm  & 3.48  & 1.0 & $\frac{\lambda_1}{10 n}$  & 10 nm  & 10 nm  & 356 nm \\ 
		\end{tabular}
		\caption{Values for the physical, material and discretization parameters used in the example in Sec.~\ref{SEC:BEAMSPLITTER}. The wavelength $\lambda$, the refractive indices for Si and air, $n_{Si}$ and $n_{BG}$, the element size used to discretize the model away from the design $h_e$ and in the design domain $h_{dx}$,  $h_{dy}$. The width of the Gaussian-envelope $\Delta y_g$.} \label{TAB:BEAMSPLITTER_OTHER}
	}
\end{table}

Parameters related to manipulating the design field and solving eq.~(\ref{EQN:OPTIMIZATION_PROBLEM_BEAM_SPLITTER}) are listed in Tab.~\ref{TAB:BEAMSPLITTER_OPTIMIZATION}.

\begin{table}[h!]
	\centering{
		\begin{tabular}{l|cccccccccc}
			Parameter & $\xi_{ini}$ & $\beta_{ini,1}$ & $\beta_{ini,2}$ & $\eta_1$ & $\eta_{2,1}$ & $\eta_{2,2}$ & $\eta_{2,3}$ & $r_{f,1}$ & $r_{f,2}$ & $n_{\mathrm{iter}}$ \\ \hline
			\\ [-1em]
			Value & 0.5 & 5 & 2.5 & 0.5 & 0.3 & 0.5 & 0.7 & 100 [nm] & 50 [nm] & 1000  \\
		\end{tabular}
		\caption{Values for quantities related to manipulating the design field and solving optimization problem in eq.~(\ref{EQN:OPTIMIZATION_PROBLEM_BEAM_SPLITTER}).} \label{TAB:BEAMSPLITTER_OPTIMIZATION}
	}
\end{table}

\noindent Given the choice of parameters the demultiplexer is designed for near-uniform erosion(dilation) of  $\pm 8$ nm around the nominal design, approximating variation that may be experienced in electron beam lithography from over(under) exposure during fabrication.

\begin{figure}[h!]
	\centering
	{
		\includegraphics[width=0.9\textwidth]{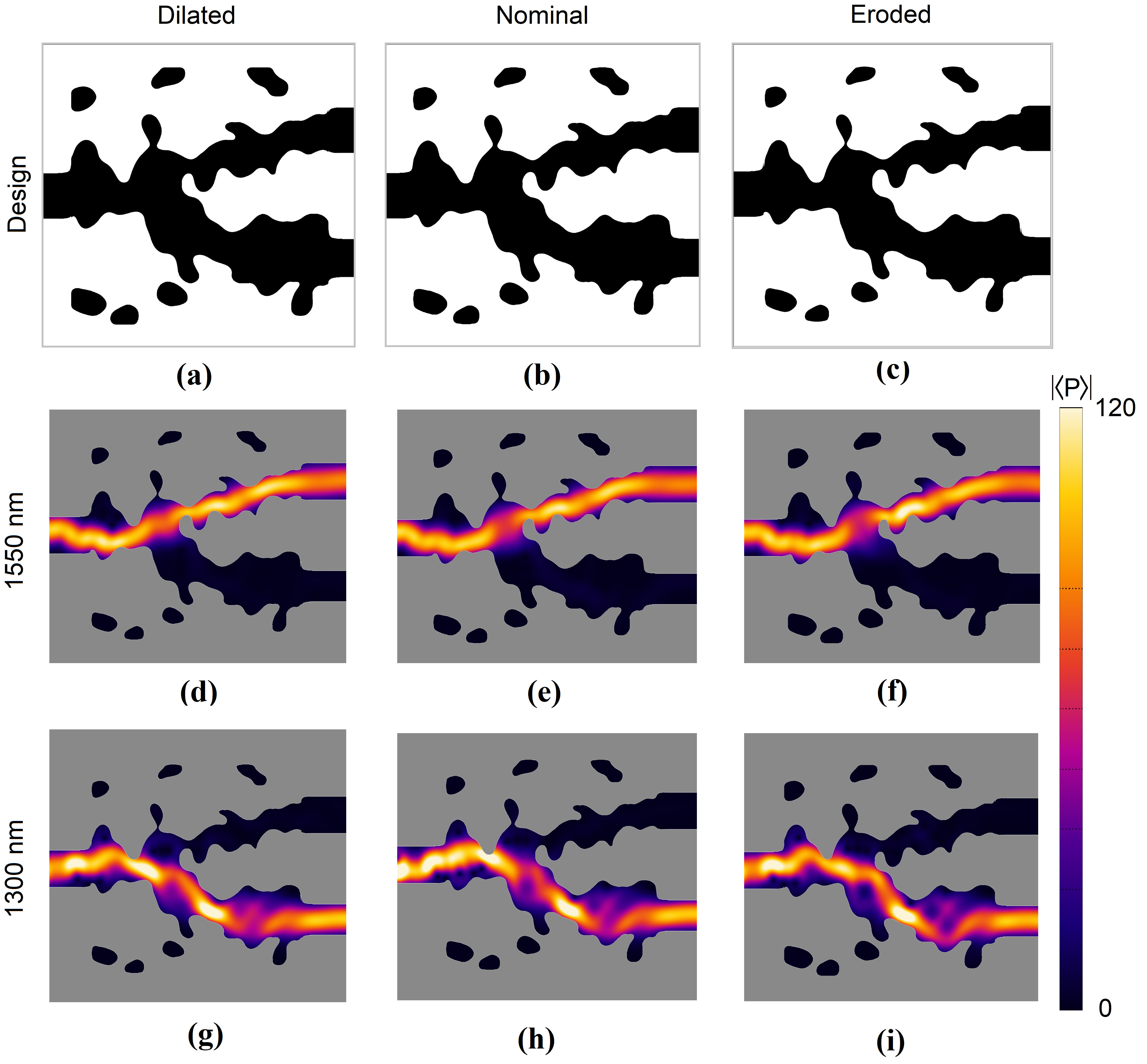} \caption{\textbf{(a)} Dilated, \textbf{(b)} Nominal and \textbf{(c)} Eroded demultiplexer design with black(white) corresponding to silicon(air). \textbf{(d-f)} Magnitude of the power flow at $\lambda = 1300$ nm for the designs in \textbf{(a-c)}.  \textbf{(g-i)} Magnitude of the power flow at $\lambda = 1550$ nm for the designs in \textbf{(a-c)}. \label{FIG:BEAM_SPLITTER_RESULTS}}
	}
\end{figure}

The optimization problem may be solved by executing the \textit{Continuation} study in the model \\ \noindent \verb|DemultiplexerExample.mph|. Doing so results in the design fields presented in Fig.~\ref{FIG:BEAM_SPLITTER_RESULTS}(a-c). The absolute value of the power flow through the demultiplexer at $\lambda= 1300$ nm is plotted in Fig.~\ref{FIG:BEAM_SPLITTER_RESULTS}(d-f) and at $\lambda = 1550$ nm in Fig.~\ref{FIG:BEAM_SPLITTER_RESULTS}(g-i).

The absolute and relative transmittance for the designs in Fig.~\ref{FIG:BEAM_SPLITTER_RESULTS}(a-c) are listed in Tab.~\ref{TAB:BEAMSPLITTER_RESULTS}. The absolute transmittance is computed as the power flow through the output waveguides ($\Gamma_1$ and $\Gamma_2$ in Fig.~\ref{FIG:BEAM_SPLITTER_MODEL_PROBLEM}) relative to the power flow through a waveguide of identical width to the input waveguide, excited identically. The relative transmittance is computed as the power flow through the output waveguides  ($\Gamma_1$ and $\Gamma_2$ in Fig.~\ref{FIG:BEAM_SPLITTER_MODEL_PROBLEM}) relative to the power flow through the input waveguide  ($\Gamma_\mathrm{in}$ in Fig.~\ref{FIG:BEAM_SPLITTER_MODEL_PROBLEM}).

The optimization problem is formulated such that identical power flow, through the relevant output waveguide, across the six cases is preferable to maximize the FOM. Looking at the first three columns in Tab.~\ref{TAB:BEAMSPLITTER_RESULTS} and scaling these by the total power flow through the reference waveguide, this is exactly what is observed for the \textcolor{black}{optimized} demultiplexer design. If desired, one could trivially change the weighting of the individual cases, in order to target larger(smaller) transmittance for a particular case.

\begin{table}[]
	\centering
	\begin{tabular}{c|cccccc}
	   & Fig.~\ref{FIG:BEAM_SPLITTER_RESULTS}(a) & Fig.~\ref{FIG:BEAM_SPLITTER_RESULTS}(b)  & Fig.~\ref{FIG:BEAM_SPLITTER_RESULTS}(c) & Fig.~\ref{FIG:BEAM_SPLITTER_RESULTS}(a) & Fig.~\ref{FIG:BEAM_SPLITTER_RESULTS}(b)  & Fig.~\ref{FIG:BEAM_SPLITTER_RESULTS}(c) \\ \hline
		\\ [-1em]
	$\lambda \ \backslash$	$T_{\bullet}$  & $T_{\mathrm{Abs}}$ & $T_{\mathrm{Abs}}$  &  $T_{\mathrm{Abs}}$  & $T_{\mathrm{Rel}}$  & $T_{\mathrm{Rel}}$  & $T_{\mathrm{Rel}}$  \\ \hline
\\ [-1em]
		 1300 nm  & $\approx 0.32$  & $\approx 0.32$  & $\approx 0.32$  & $\approx 0.86$  & $\approx 0.69$  & $\approx 0.82$   \\ \hline
		\\ [-1em]
		1550 nm& $\approx 0.31$  & $\approx 0.31$  & $\approx 0.30$  & $\approx 0.83$  & $\approx 0.85$  & $\approx 0.81$  \\ 
	\end{tabular}
	\caption{Absolute and relative transmittance for the designs in Fig.~\ref{FIG:BEAM_SPLITTER_RESULTS}. The total power flow through the reference waveguide is $\approx 7.06  \cdot 10^{-4}$ W/m at $\lambda = 1300$ nm and $\approx 7.47 \cdot 10^{-4}$ W/m at $\lambda = 1550$ nm. \label{TAB:BEAMSPLITTER_RESULTS}}
\end{table}

\section{Brief Discussions on Useful Tools} \label{SEC:ADVANCED}

Since its inception in the late 1980s a range of auxiliary tools have been developed for use with density-based Topology Optimization. While it is outside the scope of this tutorial to demonstrate all these tools, the following subset are discussed in brief.

The performance and geometry of structures designed using TopOpt have, in some cases, been found to depend strongly on the choice of the material interpolation function (e.q. eq.~(\ref{EQN:LINEAR_INTERPOLATION})). For this reason, a number of different interpolation schemes have been developed for different applications. It is thus advisable to dedicate time and effort to identifying a good interpolation scheme for a given problem. One example is the design of plasmonic nanoparticles for localized extreme field enhancement, where a non-linear interpolation scheme was demonstrated to outperform several other interpolations schemes (like eq.~(\ref{EQN:LINEAR_INTERPOLATION})) by orders of magnitude in terms of the final design performance \cite{CHRISTIANSEN_VESTER_2019}.

When designing structures for some photonic and plasmonic applications, the optimized geometries have been found to contain features with details on the order of a few nanometers \cite{WANG_2018,CHRISTIANSEN_OE_2020}. However, even with state of the art fabrication techniques, there is a lower limit to the manufacturable feature size. To ensure that designs are optimized while adhering to fabrication limitations a number of tools for imposing a minimum length-scale in the design have been developed. If optimizing for geometric robustness, length-scale may be imposed straightforwardly using the double filter technique \cite{CHRISTIANSEN_SMO_2015}\footnote{Assuming a constant topology across all design realizations.}. If the design problem is highly sensitive to geometric perturbations, making it impossible to design high performance geometrically robust structures, or if geometric robustness is not a concern for the design at hand, one may instead use a geometric constraint approach \cite{ZHOU_SMO_2015} to impose a minimum length-scale.

For some problems, such as the design of photonic membrane structures, physics dictates that all members of the structure must be connected, as free-floating members are impossible to realize. Using TopOpt it is straight forward to include a connectivity constraint in the design process, e.g. by using a virtual temperature method \cite{LI_2016}.

For some fabrication techniques, only specific design variations are allowed. As an example, in standard electron beam lithography the design blueprint must be two dimensional as little-to-no variation of the design in the out-of-plane direction is possible. Using TopOpt, it is straightforward to keep the design field constant in a particular spatial dimension by applying a simple mapping operation to the design field and integrate the design sensitivities in that spatial dimension to attain correct sensitivity information, while maintaining a constant design geometry in that direction. Other fabrication techniques allow for a smooth variations of the design in the out-of-plane direction, while simultaneously limiting these variations through a maximally allowable structural slant angle. Using TopOpt it is simple to create such designs with varying height and a limited (or fixed) slant angle using a smoothed threshold operation \cite{CHRISTIANSEN_2020b}.

Finally, when considering the use of the density-based TopOpt approach described here as a design tool for photonic structures, it is worth noting that the method has been demonstrated to be able to eliminate the need for applying proximity-effect-correction (PEC) in both electron-beam and optical-projection lithography \cite{ZHOU_ET_AL_2014} by modifying the filtering and threshold procedure and using the design field directly as the exposure dose or fabrication mask, respectively. Further the filtering and threshold procedure used in TopOpt has been demonstrated to be applicable for performing the PEC step for electron-beam lithography using an optimization based procedure \cite{ERIKSEN_ET_AL_2018}. 

\textcolor{black}{Many additional tools and techniques have been developed and explored, such as design variable linking \cite{CHRISTIANSEN_2016} and accounting for random geometric uncertainties using perturbation techniques \cite{LAZAROV_2012}. An overview of a range of different tools for ensuring lengths-scale and manufacturability may be found in \cite{LAZAROV_ET_AL_2016}.} 

\section{Conclusion}

We have presented a tutorial for applying Topology Optimization to photonic structural design, using the design of a set of cylindrical metalenses and a demultiplexer as examples of applications. First, a simple naive TopOpt problem formulation was derived and it was demonstrated that this formulation lead to several problems. Iteratively, a number of well established methods were introduced in the problem formulation and it was demonstrated how these enabled the design of physically realizable and geometrically robust structures. 

While this work  for simplicity only considers examples in two spatial dimension, an extension of the method to three spatial dimensions is trivial from the point of view of the Topology Optimization method. Using the COMSOL Multiphysics based software provided with this work, it is only a matter of using a \textit{3D component} instead of a \textit{2D component} for modelling the physics. The main challenge when extending the method from two to three spatial dimensions is the computational bottleneck associated with solving the electromagnetic state equation(s) for large-scale problems. This is however a challenge related to the numerical modelling of the physics rather than to the TopOpt method. One possible approach for treating (some) large-scale three dimensional TopOpt problems is to consider a time domain model for the physics and using finite difference time domain solvers \cite{ELSIN_ET_AL_2014} another is to use an overlapping domains techniques \cite{GANDER_ZHANG_2019,ZIN_JOHNSON_2019} to partition the physics problem into computationally tractable sub-problems. 

The reader is invited to adapt the software provided with this work to their photonics research, hereby unlocking the power of TopOpt for the design and optimization of structures for their particular applications.

\textcolor{black}{Finally, we invite the more numerically inclined reader to study our accompanying MATLAB tutorial paper \cite{CHRISTIANSEN_SIGMUND_MATLAB_2020}, which apart from a 200 line compact and transparent MATLAB implementation of TopOpt problems similar to the ones discussed here, also includes a short comparison to a non-gradient genetic-algorithm-based approach.}

\appendix

\section{COMSOL Model Description} \label{APN:COMSOL_MODEL_DESCRIPTION}

\noindent This appendix provides a brief description of the COMSOL Multiphysics model \\ \verb!MetalensCase1.mph! used to design the first iteration of a metalens in Sec.~\ref{SEC:METALENS_CASE_1}. \\

\noindent In the model the \textit{Global Definitions} contains the definitions of all model parameters defined in Tabs.~\ref{TAB:METALENS_GEOMETRY}-\ref{TAB:METALENS_OPTIMIZATION_PARAMETERS}, such as the lens width, the targeted wavelength and the design resolution.  

The framework, used to set up and model the physics and design problem, is the standard \textit{2D component}. Under the \textit{2D component}, the \textit{Definitions} node is used to define: The objective function (figure of merit); The operations related to the design field; The operations related to plotting the solutions; The material interpolation function, eq.~(\ref{EQN:LINEAR_INTERPOLATION}); A probe for printing the figure of merit; All mapping operators used to manipulate the design field; The perfectly matched layer domains. The \textit{Geometry} node, sets up all geometric elements used to build the model domain (see Fig.~\ref{FIG:META_LENS_MODEL_PROBLEM}). The \textit{Materials} node contains definitions of all the material parameters for the non-designable regions of the model domain. The \textit{Electromagnetic Waves, Frequency Domain} node defines and configures the physics model (eq.~(\ref{EQN:MAXWELL_EQUATION_E_FREQUENCY_DOMAIN})). In order to enable optimization the material parameters in the \textit{Wave Equation, Electric} sub-node is set to \textit{User defined} and the relative permittivity is set equal to the material interpolations function, eq.~(\ref{EQN:LINEAR_INTERPOLATION}). Two \textit{Scattering Boundary Condition} sub-nodes are defined, where one is used to introduce the incident field into the model domain along the lower boundary and the other is applied to the remaining boundaries of the domain. The \textit{Optimization} node is used to define the optimization problem, i.e. the figure of merit and the design variable constraint from eq.~(\ref{EQN:OPTIMIZATION_PROBLEM_GENERAL}). Finally the \textit{Mesh} node is used to setup and construct the finite element mesh for the model domain.

Two \textit{Studies} are included in the model. The first is named \textit{Optimization} and is used to execute the design procedure. In this study an \textit{Optimization} node is added to define the optimization method used, the maximum allowed number of model evaluations and the type of optimization. The \textit{Frequency Domain} study step defines the frequencies that are targeted in the optimization and the \textit{physics interfaces} that are part of the study step. Executing this study executes the optimization. The second \textit{study} is named \textit{Analysis} and is used to analyse the final design by performing a narrow band frequency sweep around the targeted frequency. 

Finally in the \textit{Results} node five \textit{Derived Values} are defined to compute the figure of merit and the power flow through the focal point and focal plane. Further, eight plots are setup to allow easy visualization of the optimized design, the resulting $\vert \textbf{E} \vert^2$-field, the power flow in the focal plane for the optimized structure obtained in the \textit{Optimization} study and the structure under analysis in the \textit{Analysis} study. 

Note that a set of auxiliary parameters, not mentioned in the paper, are defined in the COMSOL Multiphysics model(s) for practical purposes, making it easier to set up the model geometry etc.

\section*{Funding}

This work was supported in by Villum Fonden through the NATEC (NAnophotonics for TErabit Communications) Centre (grant no.~8692) and by the Danish National Research Foundation through NanoPhoton Center for Nanophotonics (grant no.~DNRF147).

\section*{Disclosures} 

The authors declare that there are no conflicts of interest related to this article.

\bibliographystyle{ieeetr} 
\bibliography{References}

\end{document}